\documentclass{ws-mplb}
\begin{document}

%\markboth{Authors' Names}{Instructions for Typing Manuscripts (Paper's Title)}

\catchline{}{}{}{}{}
%
%%%%%%%%%%%%%%%%%%%%%%%%%%%%%%%%%%%%%%%%%%%%%%%%%%%%%%%%%%%%%%%%%%%%

\title{SUPERFLUID, STAGGERED STATE, AND MOTT INSULATOR OF REPULSIVELY INTERACTING THREE-COMPONENT FERMIONIC ATOMS IN OPTICAL LATTICES
}

\author{KENSUKE INABA}

\address{NTT Basic Research Laboratories, NTT Corporation, Atsugi 243-0198, Japan\\
inaba.kensuke@lab.ntt.co.jp}
\address{JST, CREST, Chiyoda-ku, Tokyo 102-0075, Japan}

\author{SEI-ICHIRO SUGA}

\address{Department of Materials Science and Chemistry, University of Hyogo, Himeji 671-2280, Japan\\
suga@eng.u-hyogo.ac.jp}

\maketitle

\begin{history}
%\received{(Day Month Year)}
%\revised{(Day Month Year)}
\end{history}

\begin{abstract}
We review our theoretical analysis of repulsively interacting three-component fermionic atoms in optical lattices. 
We discuss quantum phase transitions at around half filling with a balanced population by focusing on Mott transitions, staggered ordering, and superfluidity.
At half filling (with 3/2 atoms per site), characteristic Mott transitions are induced by the anisotropic interactions, where two-particle repulsions between any two of the three colors have different strengths.
At half filling, two types of staggered ordered states appear at low temperatures depending on the anisotropy of the interactions. 
As the temperature increases,  phase transitions occur from the staggered ordered states to the unordered Mott states. 
Deviating from half filling, an exotic superfluid state appears close to a regime in which the Mott transition occurs. 
We explain the origin of these phase transitions and present the finite-temperature phase diagrams. 
\end{abstract}

\keywords{three-component fermions; superfluid; paired Mott insulator; color-selective Mott transition.}

%%%%%%%%%%%%%%%%%%%%%%%%%%%%%%%%%%%%%%%%%%%%%%
\section{Introduction}
%%%%%%%%%%%%%%%%%%%%%%%%%%%%%%%%%%%%%%%%%%%%%%

The ultracold atoms in optical lattices provide a new way for studying quantum many-body phenomena, which are the essence of solid state physics.\cite{Bloch,Bloch2005,Jaksch2005,Morsch2005,Greiner2008} 
The high tunablility of artificial lattices and atomic interactions allows us to simulate quantum phase transitions. 
In fact, the Mott transition has been experimentally observed in a wide variety of settings.\cite{Chin2006,Jordens2008,Schneider2008,Fukuhara2009,Greiner2002,Gunter2006,Gadway2010}
The high controllability of atoms further provides novel many-body systems beyond those found in solids, thereby extending our insights. 
A prominent example is Bose-Fermi mixtures in optical lattices, in which a characteristic Mott insulating state has been observed.\cite{Sugawa2011,Gunter2006,Ospelkaus2006}
As a result, the  Mott transition paradigm has been generalized to mixtures of particles with different quantum statistics. 
Another example is fermionic atoms with multiple internal degrees of freedom. 
In particular, odd-number-component systems have noteworthy features that distinguish them from well-studied even-number-component systems.
Quantum degenerate gases of three-component fermionic $^6$Li atoms have been created in experiments.\cite{Ottenstein2008,Huckans2009} 
Theoretically, it has been clarified that three-component atoms trapped in optical lattices undergo characteristic quantum phase transitions.\cite{HH2004B,Rapp2007,Demler2007,Rapp2008,Inaba2009a,Klings2010,Inaba2011,Priv,Titv,Rapp2012,Miyatake2010,Inaba2010b,Suga2011,Inaba2012,Suga2012,Gorelik,Caponi2008,Molina2009,Azaria2009,Kantian2009,Rapp2011}
Other fascinating systems are fermionic atoms with exotic symmetries, such as ${\rm ^{173}Yb}$ atoms with SU(6) symmetry,\cite{Fukuhara2007,Taie2012} mixtures of ${\rm ^{173}Yb}$ and ${\rm ^{171}Yb}$ atoms with SU(6)$\times$SU(2) symmetry,\cite{Taie2010} and ${\rm ^{87}Sr}$ atoms with SU(10) symmetry.\cite{DeSalvo2010} 
Theoretical analyses have shown that SU($N$) fermions with a large $N$ exhibit novel ordered states.\cite{AM1988,MA1989,HH2004a,Hermele2009,Cazalilla2009,Gorshkov2010,Yip2011,Blumer} 
In the meantime, one of the most attractive research fields in current condensed-matter physics is multi-component systems, which correspond to strongly correlated electrons with orbital degeneracy.
For instance, the origin of  iron-based superconductors can be understood by considering a role of  orbital degeneracy.\cite{Stewart2011,Mazin2011}
Multi-component cold-atom systems thus have the potential to contribute to great development of quantum many-body physics.

In this review, we discuss the quantum phase transitions of repulsively interacting three-component (color degrees of freedom) fermionic atoms in optical lattices.\cite{Miyatake2010,Inaba2010b,Suga2011,Inaba2012,Suga2012}  
We have numerically analyzed the fermionic Hubbard model using the dynamical mean-field theory (DMFT).\cite{Georges1996}
In particular, we focus on the model close to half filling with a balanced population of three colors, and we consider the effects of anisotropic (color-dependent) interaction strengths. 
The balanced population and the anisotropic interactions are realistic in $^6$Li systems.\cite{Ottenstein2008}
We have found that anisotropic interactions induce characteristic Mott transitions at half filling (with 3/2 atoms per site). 
This is remarkable for non-integer number of atoms per site, and we have clarified the origin of these unusual Mott transitions.\cite{Inaba2010b}  
We have also investigated staggered-ordered ground states,\cite{Miyatake2010} extending a prior work\cite{Gorelik}, and examined the phase transitions between the ordered states and the Mott states that occur at finite temperatures.\cite{Suga2011}
%
%Our findings regarding the Mott transition added to our knowledge of the staggered ordered state discussed in a previous study.\cite{HH2004a} 
%
Furthermore, we have found that an exotic $s$-wave superfluid state appears close to the characteristic Mott phase, although atoms are interacting repulsively  with each other.\cite{Inaba2012,Suga2012}
The mechanism of this superfluid can be understood in relation to the origin of the Mott transition.\cite{Miyatake2010,Inaba2010b,Suga2011,Inaba2012,Suga2012}

The rest of this paper is organized as follows. 
Section \ref{sec_Model} describes the model Hamiltonian and our theoretical methods.
Section \ref{sec_Mott} reviews the nature of the characteristic Mott transitions at half filling and also provides an overview of the other phase transitions.
Section \ref{sec_Mag} discusses the properties of the staggered ordered states and presents a finite-temperature phase diagram at half filling.
Section \ref{sec_SF} analyzes the appearance of the $s$-wave superfluid state close to half filling and clarifies the mechanism of the Cooper pairing that induces the superfluid.
Section \ref{sec_Summary} provides a summary and an outlook.

%%%%%%%%%%%%%%%%%%%%%%%%%%%%%%%%%%%%%%%%%%%%%%
\section{Model and Methods}\label{sec_Model}
%%%%%%%%%%%%%%%%%%%%%%%%%%%%%%%%%%%%%%%%%%%%%%
In this section, we introduce our model and briefly explain the numerical methods used in our studies. 
The low-energy properties of the ultracold atoms in an optical lattice are well described by the following Hubbard Hamiltonian:\cite{Jaksch} 
%""""""""""""""""""""""""""""""""""""
\begin{eqnarray}
\hat{\cal H}=-t \sum_{\langle i,j \rangle}\sum_{\alpha=1}^{3}
       \hat{a}^\dag_{i\alpha} \hat{a}_{j\alpha} 
  - \sum_{i}\sum_{\alpha=1}^{3} \mu_\alpha \hat{n}_{i \alpha} 
  + \frac{1}{2}\sum_{i}\sum_{\alpha\not=\beta} 
       U_{\alpha\beta} \hat{n}_{i \alpha} \hat{n}_{i \beta},   
\label{eq_model}
\end{eqnarray}
%____________________________________
where $U_{\alpha\beta}$ denotes the onsite interactions between color-$\alpha$ and $\beta$ atoms, $t$ is the nearest-neighbor hopping integral, and $\hat{a}^\dag_{i\alpha} (\hat{a}_{i\alpha})$ and $\hat{n}_{i\alpha}$ are creation (annihilation) and number operators of a fermion with color $\alpha$ at the $i$th site. 
The subscript $\langle i,j \rangle$ is the summation over the nearest-neighbor sites.
Filling $N$ is given by $N=\sum_\alpha n_\alpha$, where $n_\alpha\equiv \langle \hat{n}_{i \alpha} \rangle$ is the average number of color-$\alpha$ atoms at a site.
This paper focuses only on the situation where the populations of different colors are balanced, $n\equiv n_1=n_2=n_3$.
The populations can be controlled by tuning the chemical potentials $\mu_\alpha$.
For half filling ($N\equiv 3n=3/2$), the particle-hole symmetry yields $\mu_\alpha=(U_{\alpha\beta}+U_{\alpha\gamma})/2$. 
To investigate the essential features of the correlation effects, the lattice geometry discussed here is restricted to the infinite-dimensional Bethe lattice without the trapping potential.
Non-interacting atoms in this lattice have a semicircular density of states given by $\rho_0(\omega)=\sqrt{4t^2-\omega^2}/(2\pi t^2)$. 
In what follows, the hopping integral $t$ is used as the unit of energy, and the Planck constant and the Boltzmann constant are set to unity for simplicity: $\hbar=k_B=1$.
%For convenience, $U_{\alpha\beta}$ are denoted by $U_{12}\equiv U$, $U_{23}\equiv U'$, and $U_{31}\equiv U''$. 
The realization of this model is discussed in Sec. \ref{sec_Summary}.

Let us briefly explain the DMFT, which is the theoretical framework employed in our studies. 
The DMFT allows us to deal exactly  with  local correlations in infinite dimensional systems, and thus to capture the essential nature of the Mott transition.\cite{Georges1996}
The DMFT can also be used to investigate the $s$-wave superfluidity in the Hubbard model,\cite{Garg2005,Miyatake2010,Inaba2010b,Inaba2012} because its order parameter is described by the onsite correlations $\langle \hat{a}_{\alpha i} \hat{a}_{\beta i} \rangle$. 
Since the Bethe lattice is bipartite, the staggered ordered states are expected to be stable at low temperatures. 
Such states can be investigated by extending the DMFT to the two-sublattice treatment.\cite{Georges1996}
Thus, this method can also capture the essential properties of the phase transitions among the staggered ordered phases and the superfluid phase.
%

%%%%%%%%%%%%%%%%%%%%%%%%%%%%%%%%%%%%%%%%%%%
\section{Mott transitions}\label{sec_Mott}
%%%%%%%%%%%%%%%%%%%%%%%%%%%%%%%%%%%%%%%%%%%
In this section, we discuss the Mott transition specific to  repulsively interacting three-component fermions at half filling with a balanced population.
At  first glance, we expect the Mott transition to disappear at this filling, because the number of atoms in each lattice site is non-integer; $N=3/2$. 
In fact,  previous studies have shown that no Mott transition occurs at the SU(3) point with isotropic (color-independent) interactions $U_{12}=U_{23}=U_{31}$.\cite{Gorelik} 
However, we found that characteristic Mott transitions are induced by the anisotropy of the interactions $U_{12}\not=U_{23}\not=U_{31}$.\cite{Miyatake2010,Inaba2010b} 
As  discussed later in this section, there are two types of  Mott states: a color-selective Mott state (CSM) and a paired Mott insulator (PMI).

\begin{figure}[tb]
\begin{center}
\includegraphics[scale=0.55]{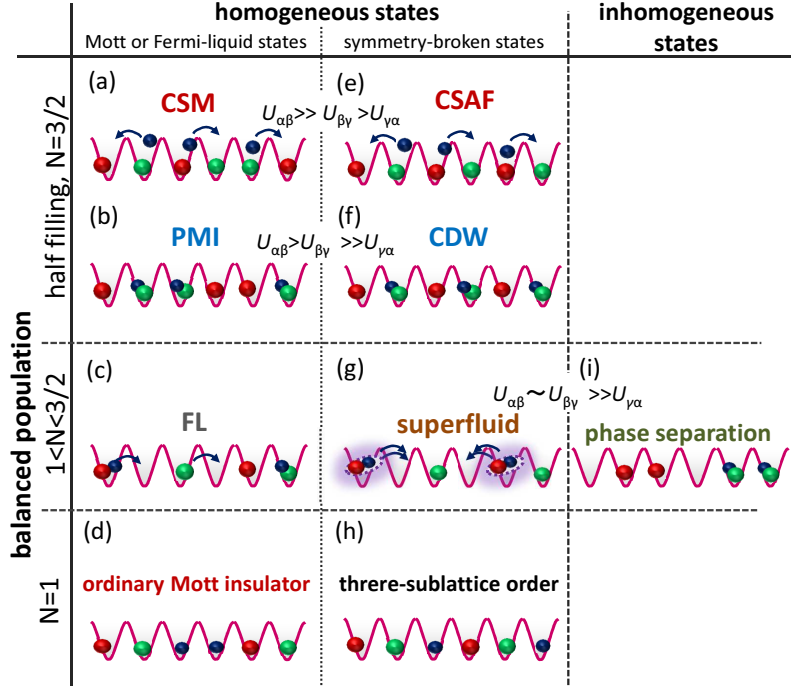}
\caption{
Schematic pictures of possible states in the system with a balanced population for various fillings $N$. 
(a)-(d): Homogeneous states without symmetry broken, (e)-(h): those with symmetry broken, and (i): inhomogeneous state. 
This figure is devoted to a guide to overviewing the results shown in Secs. \ref{sec_Mott}, \ref{sec_Mag}, and \ref{sec_SF}. 
%The present study is focused on the system equal and close to half filling $N=3/2$. 
%Because of particle-hole symmetry, the states in $N}3/2$ are excluded for simplicity. We specify three internal degrees of freedom ($\alpha$, $\beta$, $\gamma$) by three colors (red, green, blue), respectively. 
At half filling $N=3/2$, (a) color-selective Mott state (CSM) and (b) paired Mott insulator (PMI) are found when one of the interactions is strong ($U_{\alpha\beta}\gg U_{\beta\gamma}, U_{\gamma\alpha}, t$) and two of the interactions are strong ($U_{\alpha\beta}, U_{\beta\gamma}\gg U_{\gamma\alpha}, t$), respectively (see Sec. \ref{sec_Mott}). Their corresponding staggered ordered states are (e) color-selective antiferromagnet (CSAF) and (f) color-density wave (CDW), respectively (see Sec. \ref{sec_Mag}). 
For $1 \le N \le 3/2$, (c) Fermi liquid (FL) appears. Close to (but not at) $N=3/2$, (g) superfluid appears in moderately correlated region with $U_{\alpha\beta}\ll U_{\beta\gamma}, U_{\gamma\alpha}$ (see Sec. \ref{sec_SF}). In further strongly correlated region, (i) phase separation occurs (see Sec. \ref{sec_SF}).
The states at integer filling $N=1$ are shown for comparison. 
Here, (d) ordinary Mott insulator\protect\cite{Gorelik} and (h) its corresponding three-sublattice ordering appear.\protect\cite{ThreeSub} 
The ordered states without the balanced population are discussed in Ref. 37. %\protect\cite{Rapp2011}.
Possible states appearing at the blank spaces has not been discussed so far.
}
\label{fig_SF}
\end{center}
\end{figure}

It should be noted that the Mott transition mechanism plays a key role in understanding the nature of subsequent staggered ordering and superfluidity. 
To provide an overview of these quantum phase transitions, we first focus on the properties of the Mott transition at zero temperature, $T=0$, by neglecting possible ordered states. 
The staggered orders and the superfluid are discussed in detail in Secs. \ref{sec_Mag} and \ref{sec_SF}, respectively.
In Fig. \ref{fig_SF}, we show schematic pictures of possible states discussed in this article. This figure is dedicated to a guide to the main results presented in Secs. \ref{sec_Mott}, \ref{sec_Mag}, and \ref{sec_SF}. 
Details are discussed in the corresponding sections.

%---------------------------------------
%---------------------------------------

\subsection{Paired Mott insulator and color-selective Mott state}\label{subsec_app_Mott}
%
%---------------------------------------
\begin{figure}[tb]
\begin{center}
\includegraphics[scale=0.4]{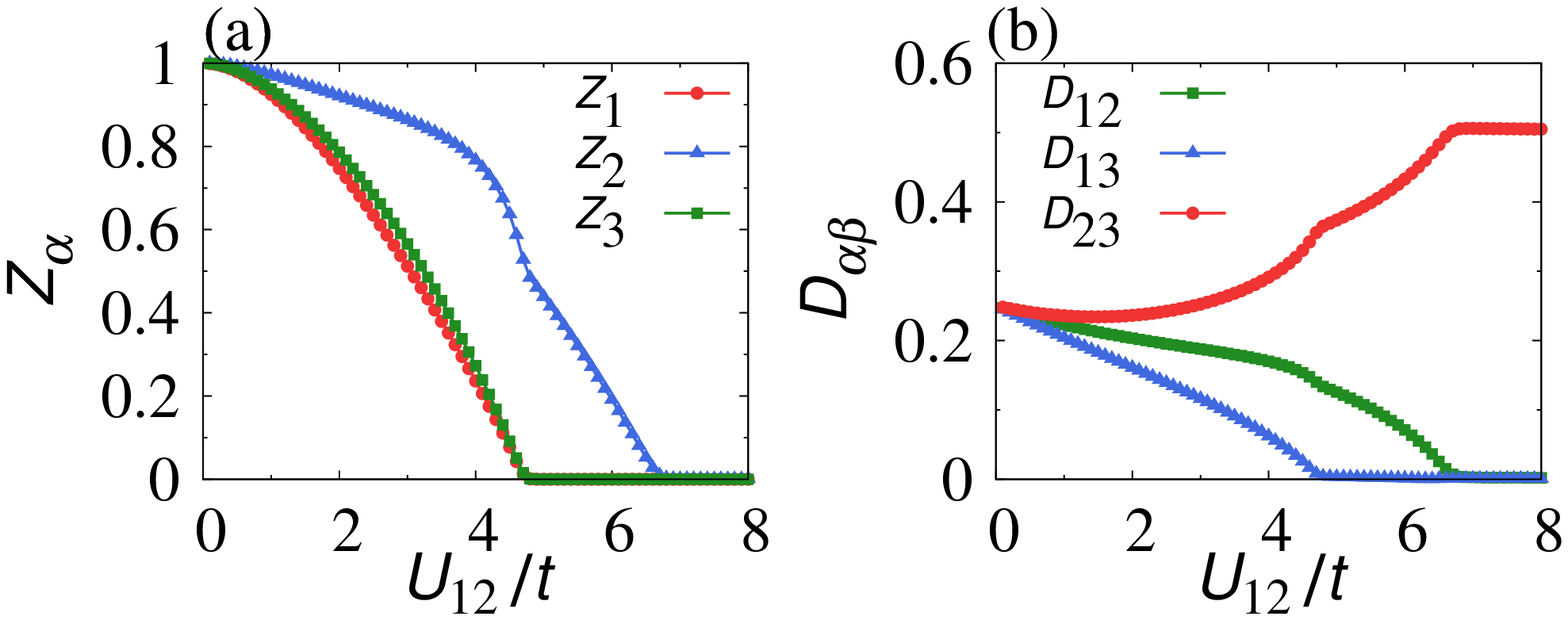}
\caption{Renormalization factors $Z_{\alpha}$ and double occupancies $D_{\alpha \beta}$ for $U_{31}/U_{12} = 1.5$ and $U_{23}/U_{12} = 0.7$ at $T=0$ as functions of $U_{12}/t$.\protect\cite{Inaba2010b}
}
\label{fig_DZvsU}
\end{center}
\end{figure}
%---------------------------------------
%
We used two-site DMFT method\cite{Potthoff2001} to calculate two quantities that characterize the Mott transition: the renormalization factors $Z_\alpha\equiv 1/(1-\partial \Sigma_\alpha(\omega)/\partial\omega)|_{\omega \to 0}$ and the double occupancies $D_{\alpha\beta}\equiv \langle \hat{n}_{i\alpha} \hat{n}_{i\beta} \rangle$, where $\Sigma_\alpha(\omega)$ is the self-energy of color-$\alpha$ atoms. 
Here, we set the interaction anisotropy at $U_{31}/U_{12} = 1.5$ and $U_{23}/U_{12} = 0.7$. 
Figure \ref{fig_DZvsU} shows $Z_\alpha$ and $D_{\alpha\beta}$ as functions of $U_{12}/t$.
The renormalization factors $Z_{\alpha}$ decrease monotonically as the interaction strength $U_{12}/t$ increases, meaning that the effective mass of the atom increases. 
The disappearance of $Z_1$ and $Z_3$ at $U_{c1}/t=4.7$ ($Z_2$ at $U_{c2}=6.7$) indicates that color-1 and 3 (color-2) atoms are localized at this point. 
As $U_{12}/t$ increases, $D_{13}$ decreases and vanishes at $U_{c1}$, and $D_{12}$ disappears at $U_{c2}$ in a similar way. 
In contrast, $D_{23}$ increases towards a saturated value $1/2$ at $U_{c2}$.  These results reveal that the metallic Fermi liquid (FL) state appears for $U_{12}<U_{c1}$, and two characteristic Mott states appear for $U_{c1}<U_{12}<U_{c2}$ and $U_{c2}<U_{12}$. 
These two Mott states, termed the color-selective Mott state (CSM) and the paired Mott insulator (PMI), are illustrated schematically in Fig. \ref{fig_SF}(a) and (b), and in comparison with an ordinary Mott state at the integer filling ($N=1$) in Fig. \ref{fig_SF}(d).
The CSM and the PMI are discussed in detail below.

{\it \bf Color-selective Mott state (CSM).} 
In Fig. \ref{fig_DZvsU}, we find that $Z_1=Z_3=0$ and $Z_2\not=0$ for $U_{c1}<U_{12}<U_{c2}$. 
This indicates that the color-1 and 3 atoms selectively undergo the Mott transition to avoid the strongest interaction $U_{31}$ between the two colors. 
We call this state CSM, which is shown schematically  in Fig. \ref{fig_SF}(a). 
Color-1 and 3 atoms are localized randomly at different sites, while color-2 atoms are itinerant throughout the system. 
This characteristic Mott transition results from the following specific commensurability: the filling of selected colors is unity, $n_{1}+n_{3}=1$. 
The itinerant color-2 atoms in the CSM can be effectively described by the Falikov-Kimball model,\cite{Miyatake2010} which is a well known toy model for $f$-electron systems.\cite{FK,Brandt}
It has been pointed out that non-FL properties occur in this model.\cite{Si1992} Similar results were reported for the orbital-selective Mott transition in the solid state context.\cite{Biermann2005} 
In Sec. \ref{subsec_Mag3}, we discuss that itinerant atoms in the CSM do exhibit non-FL behavior.

{\it \bf Paired Mott insulator (PMI).} 
For $U_{12}>U_{c2}$ in Fig. \ref{fig_DZvsU}, we find that $Z_\alpha=0$ for all $\alpha$, meaning that all the atoms are localized. 
We also find  constant double occupancies with $D_{12}=D_{13}=0$ and $D_{23}=1/2$.
This indicates that color-2 and 3 atoms form effective pairs to avoid stronger two interactions $U_{12}$ and $U_{31}$, and are thus localized at the same sites, while color-1 atoms are solely localized at the other sites.
Here, the pairs and the color-1 atoms are distributed randomly.
We call this state PMI and depict it schematically in Fig. \ref{fig_SF}(b). 
Because of the pair formation, the effective filling becomes unity as $N_{\rm eff}=n_{\rm pair}+n_{3}=1$, where $n_{\rm pair}$ is the average number of pairs at a site. 
This is  why the Mott transition occurs at the non-integer half filling. 
We note that this mechanism is analogous to the mixture of bosons and fermions, where the Mott transition has been observed when $n_{\rm boson}+n_{\rm fermon}$ becomes an integer.\cite{Sugawa2011}

\subsection{$U_{12}$-$U_{23}$ Phase diagrams}
\begin{figure}[tb]
\begin{center}
\includegraphics[scale=0.5]{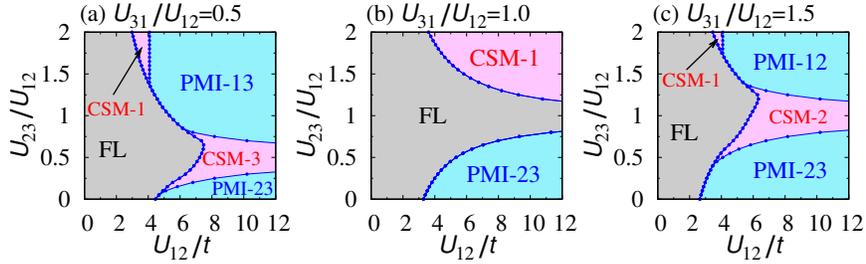}
\caption{$U_{12}$-$U_{23}$ phase diagrams of Mott states at half filling obtained for fixed $U_{31}/U_{12}$ at $T=0$.\protect\cite{Inaba2010b}
}
\label{fig_RU_phase}
\end{center}
\end{figure}

We performed the same calculations as the above while systematically changing $U_{23}/U_{12}$ for several choices of $U_{31}/U_{12}$.
We then determined the $U_{23}$-$U_{12}$ phase diagrams (Fig. \ref{fig_RU_phase}).
The above scenario for the anisotropy parameters $U_{23}/U_{12}$ and $U_{31}/U_{12}$ can be straightforwardly adapted to other combinations of $U_{12}$, $U_{23}$, and $U_{31}$. 
Here, CSM-$\alpha$ indicates that the itinerant atoms are of color-$\alpha$, and PMI-$\alpha\beta$ indicates that color-$\alpha$ and $\beta$ atoms form pairs.
We can see that there are no Mott transitions in the isotropically interacting SU(3) system (the line  $U_{23}/U_{12}=1$ in Fig. \ref{fig_RU_phase}(b)), as was originally pointed out by Gorelik and Bl\"{u}mer.\cite{Gorelik} 

%---------------------------------------
%---------------------------------------

%
\subsection{Discussions of possible ordered states}\label{subsec_discuss_Mott}
We close this section by briefly discussing possible symmetry-broken phases. 
The above calculations neglected the appearance of such phases and focused only on Mott transitions for simplicity.
However, from the characteristics of the Mott states mentioned above, we can infer their potential instabilities towards staggered ordered states and a superfluid state.

First, we discuss the possibility of staggered ordering, which corresponds to  {\it color magnetism}. 
Localized ``particles" in the Mott states are randomly distributed at different sites, and such states are energetically degenerate. 
Localized ``particles" in the CSM are atoms of two selected colors [Fig. \ref{fig_SF}(a)]; 
in the limit $U_{\alpha\beta}/t\to \infty$, the energy of a site with a localized color-$\alpha$ or $\beta$ atom is given respectively by $1/2(-2\mu_\alpha-\mu_\gamma+U_{\gamma\alpha})$ and $1/2(-2\mu_\beta-\mu_\gamma+U_{\beta\gamma})$, which are both reduced to $-1/2(2U_{\alpha\beta}+U_{\beta\gamma}+U_{\gamma\alpha})$ because of the condition $\mu_\alpha=1/2(U_{\alpha\beta}+U_{\alpha\gamma})$ at half filling. 
On the other hand, localized ``particles" in the PMI are paired atoms, which are composite particles, and third-color atoms [Fig. \ref{fig_SF} (b)]; in the limit $U_{\alpha\beta}/t\to \infty$, their energies are $-\mu_\beta$ and $-\mu_\alpha-\mu_\gamma +U_{\gamma\alpha}$, which are both reduced to $-1/2(U_{\beta\gamma}+U_{\alpha\beta})$ at half filling. 
These two degenerate states in the CSM and the PMI can be regarded as the $S=1/2$ free pseudospins, which result in the macroscopic $2^L$-fold degeneracy with $L$ being the number of lattice sites. 
These features are analogous to those of the well-known two-component ($S=1/2$ spin) Hubbard model. 
If we consider spatial fluctuations induced by the second-order perturbative hopping processes, we expect the pseudospins to form a certain long-range order, such as a staggered order, and thus the macroscopic degeneracy is lifted at low temperatures.
We discuss such ordering transitions at finite temperatures in Sec. \ref{sec_Mag}.

The color magnetism is sensitive to the filling and the population balance. 
With the balanced population, the magnetic orders become unstable when the filling deviates from half filling.\cite{Rapp2011}
As shown in Fig. \ref{fig_SF}(d), the ordinary Mott insulator appears at the filling $N=1$. 
In this situation, the effective spin model is given by the SU(3) spin Hamiltonian, and the possible ordered state is three-sublattice ordering [Fig. \ref{fig_SF}(g)] \cite{ThreeSub}. 
In addition, away from the balanced population, we can expect different types of ordered states.\cite{Rapp2011}
These ordered states are beyond of the scope of this review.

Next, we remark on the possible appearance of superfluidity.
The $U_{12}$-$U_{23}$ phase diagrams (Fig. \ref{fig_RU_phase}) in fact suggests the appearance of the superfluid. 
For the anisotropic interactions $U_{12}\not=U_{23}\not=U_{31}$, we can generally find two successive transitions, namely, the FL-CSM and CSM-PMI transitions as in Fig. \ref{fig_DZvsU}. In addition, we find a direct transition from the FL to the PMI in some specific parameter regions, where one of the three interactions is much smaller than the others, {\it e.g.,} $U_{23}\ll U_{12}\sim U_{31}$.
An example is the small-$U_{23}/U_{12}$ region of the phase diagram shown in Fig. \ref{fig_RU_phase}(b). 
In the FL phase close to the PMI, we expect the pair formation to induce strong fluctuations that can cause a superfluid transition. 
We show the schematic picture of a possible superfluid state in Fig. \ref{fig_SF}(g).
Note that the superfluid transition cannot be induced in the CSM phase because two of the three colors are localized.
The issue of superfluidity is discussed in Sec. \ref{sec_SF}. 
For further strongly correlated region in $U_{23}\ll U_{12}\sim U_{31}$, the pair formation may cause an instability towards an inhomogeneous phase as shown in Fig. \ref{fig_SF}(i). We discuss this issue in Sec. \ref{sec_SF}.

%%%%%%%%%%%%%%%%%%%%%%%%%%%%%%%%%%%%%%%%%%%%%%
\section{Staggered states}\label{sec_Mag}
%%%%%%%%%%%%%%%%%%%%%%%%%%%%%%%%%%%%%%%%%%%%%%
In this section, we discuss the phase transitions to the staggered ordered states at half filling with a balanced population.
First, let us mention the pioneering work undertaken by Honerkamp and Hofstetter. 
They pointed out that a characteristic staggered ordered state called the color-density wave (CDW) state appears for the isotropic SU(3) interactions in the square lattice at zero temperature.\cite{HH2004a}
Since the square lattice is a bipartite lattice, the perfect nesting condition with the nesting wavevector $(\pi, \pi)$ is satisfied at half filling.
We therefore expect the staggered ordered state to be stable at low temperatures. 
Let us also mention  previous studies of one-dimensional three-component lattice systems, where the nature of the staggered ordering is quite different from that of higher-dimensional systems because strong quantum fluctuations play an essential role. 
The specialized analytical and numerical techniques employed for one-dimensional systems, such as the bosonization and the density-matrix renormalization group, have revealed the occurrences of characteristic orderings.\cite{Caponi2008,Molina2009,Azaria2009,Kantian2009}

Our research focused on the infinite-dimensional Bethe lattice. 
We have found that the CDW state appears also for anisotropic interactions at zero temperature.
Changing the anisotropy systematically, we have found another staggered ordered state: a color-selective antiferromagnet (CSAF).  
 We have also revealed that the CDW and the CSAF states are both degenerate at the isotropic SU(3) point within our numerical accuracy.
Furthermore, we examined the phase transitions at finite temperatures and finally determined the $U_{12}$-$T$ phase diagrams at half filling, which include the Mott states and the staggered ordered states.

\subsection{Staggered ordered states}\label{subsec_app_Mag}
%---------------------------------------
\begin{figure}[tb]
\begin{center}
\includegraphics[scale=0.35]{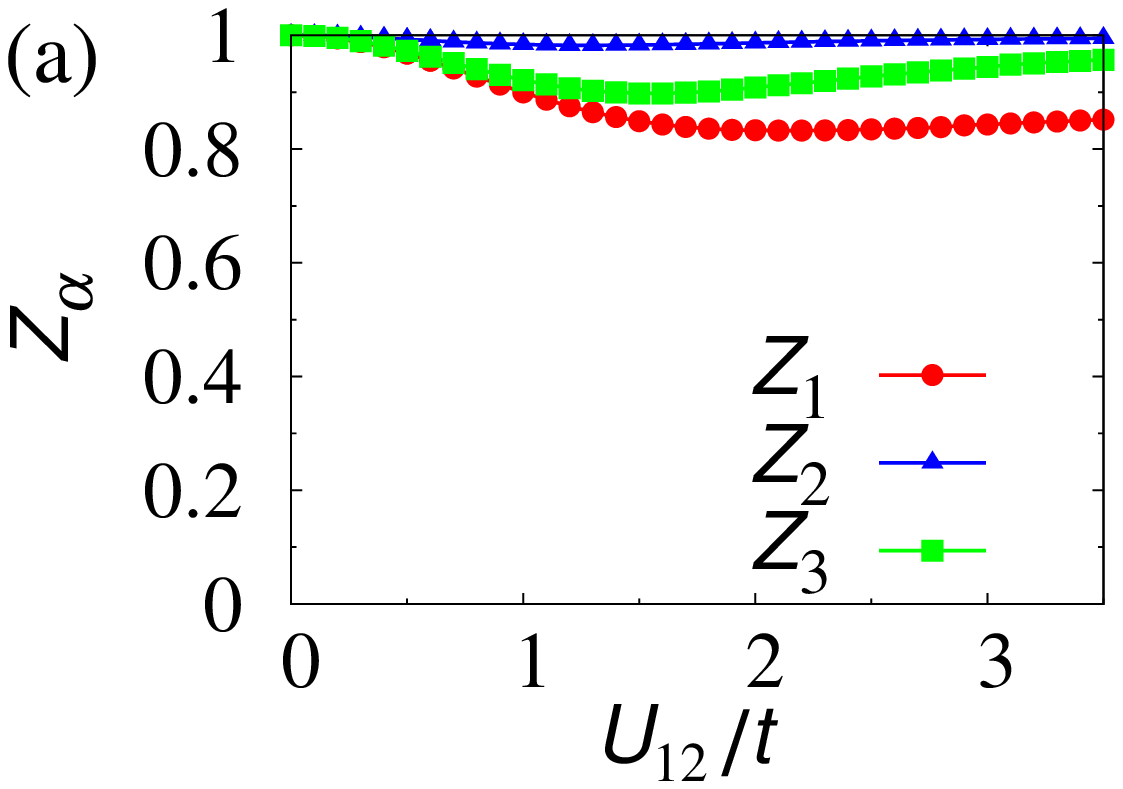}
\includegraphics[scale=0.35]{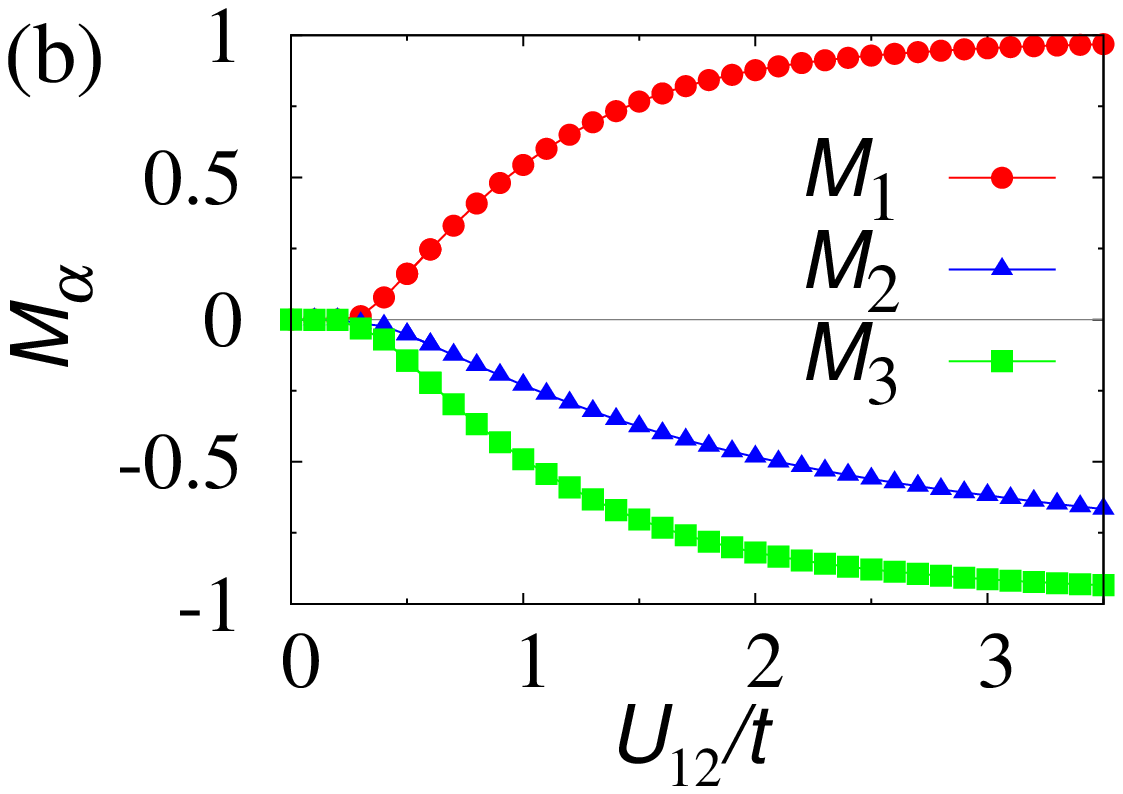}
\includegraphics[scale=0.35]{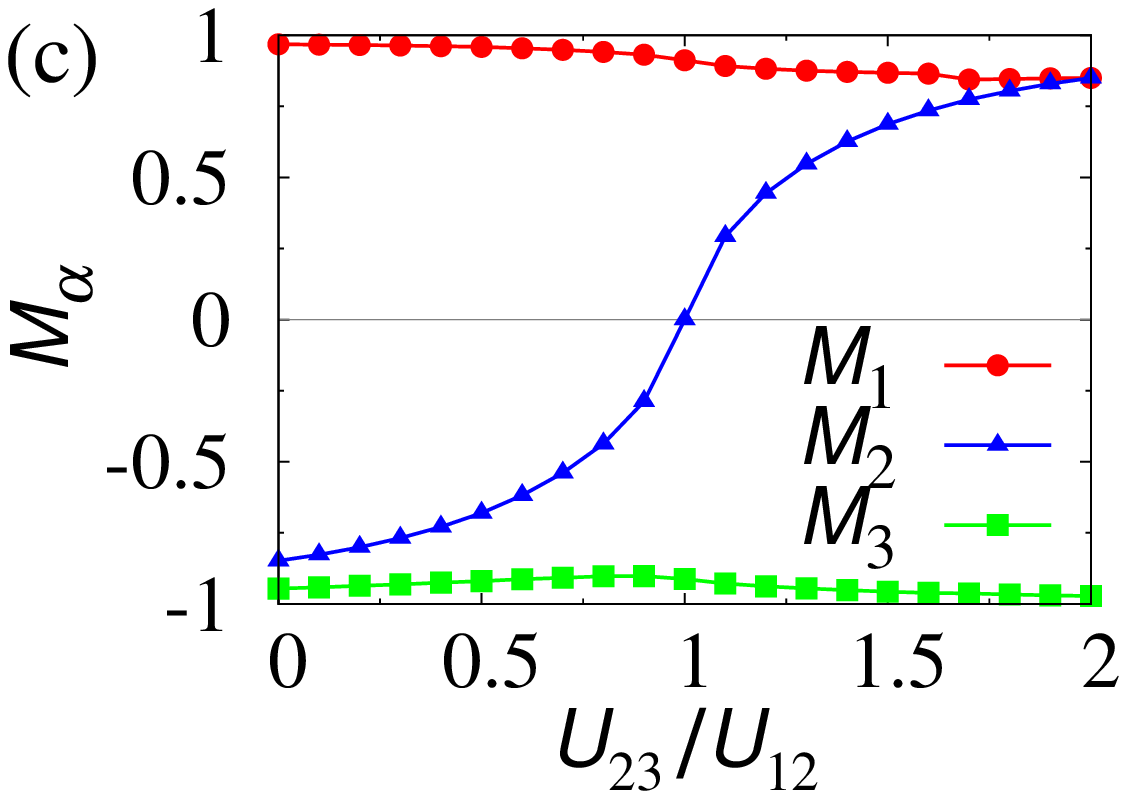}
\caption{
(a) Renormalization factors $Z_\alpha$ and (b) staggered order parameters $M_\alpha$ as functions of $U_{12}/t$ for $U_{31}/U_{12}=2$ and $U_{23}/U_{12}=0.6$.\protect\cite{Miyatake2010} 
(c) $M_\alpha$ vs. $U_{23}/U_{12}$ for $U_{31}/U_{12}=2$ at $U_{12}/t=3$.\protect\cite{Miyatake2010} 
Calculations are for half filling at $T=0$. 
}
\label{fig_MvsU}
\end{center}
\end{figure}
%---------------------------------------

To investigate the staggered ordered states, we employed the extended two-site DMFT,\cite{Miyatake2010} where the bipartite lattice is divided into two sublattices labeled $A$ and $B$. 
We calculated the staggered order parameters $M_{\alpha}=n^A_{\alpha}-n^B_{\alpha}$ and the renormalization factors $Z_{\alpha}$, where $n^{A(B)}_\alpha=\langle \hat{n}_{i\alpha} \rangle$ is defined for $i \in$ $A (B)$ sub-lattices.

Figure \ref{fig_MvsU}(a) and (b), respectively, show $Z_\alpha$ and $M_\alpha$ as functions of $U_{12}/t$ for $U_{31}/U_{12}=2$ and $U_{23}/U_{12}=0.6$ at half filling. 
As $U_{12}/t$ increases, $Z_{1}, Z_{2}$, and $Z_{3}$ first decrease and then increase without vanishing even for a large $U_{12}/t$. 
We also find that for $U_{12}/t>0.4$ $,|M_\alpha|$ gradually increase as $U_{12}/t$ increases. 
These features suggest that the staggered ordered state does appear, while the Mott transition does not occur.

We further investigated the properties of the staggered ordered states by changing $U_{23}/U_{12}$.
In Fig. \ref{fig_MvsU}(c), we show $M_{\alpha}$ as functions of $U_{23}/U_{12}$ for $U_{31}/U_{12}=2$ and $U_{12}/t=3$. 
We find that $M_{2}$ changes from negative to positive as $U_{23}/U_{12}$ increases and that $M_{1} \sim 1$ and $M_{3} \sim -1$ irrespective of $U_{23}/U_{12}$. 
These results indicate that two types of  staggered order, the CSAF and the CDW (see below), appear depending on the anisotropy of the repulsions.

{\it \bf Color-selective antiferromagnet (CSAF).} 
At $U_{23}/U_{12}=1$ in Fig. \ref{fig_MvsU}(c), the finite order parameters, $M_1 \sim 1$ and $M_3 \sim -1$, indicate that the color-1 and 3 atoms alternately occupy  different sites.
In contrast, vanishing $M_{2}$ suggests that the color-2 atoms are itinerant.
This ordered state is a CSAF, which is shown schematically in Fig. \ref{fig_SF}(e). 
The CSAF is found in a specific parameter region, where two of the three interactions are equal and smaller than the other, {\it e.g.}, $U_{31} > U_{23}=U_{12}$.
At around $U_{23}/U_{12}\sim1$, the nearly vanishing $M_2$ means that the staggered state with the localized color-2 atoms is unstable.
Accordingly, if the perfect nesting condition is violated by, for example, geometrical frustration, we expect the color-2 atoms to become itinerant in a wide region around $U_{23}/U_{12}=1$ and the CSAF appears there.

{\it \bf Color density wave (CDW).}  
From Fig. \ref{fig_MvsU}(c), we find positive $M_{1}$ and  negative $M_{2}$ and $M_3$ for $U_{23}/U_{12}<1$, while we find positive $M_1$ and $M_2$ and negative $M_3$ for $U_{23}/U_{12}>1$.
These results indicate that color-2 and 3 (color-1 and 2) atoms occupy the same sites, and the adjacent sites are solely occupied by color-1 (3) atoms for $U_{23}/U_{12}<1$ ($U_{23}/U_{12}>1$).
In other words, paired atoms with two different colors and atoms with the third color alternately occupy different sites. 
Figure \ref{fig_SF}(f) depicts this characteristic staggered ordered state, namely, a CDW.

Here, we explain the relationship between the CSAF and the CSM, and that between the CDW and the PMI.
The staggered ordered states (CSAF and CDW) can be regarded as  antiferromagnetic states formed by the $S=1/2$ pseudospins in the corresponding Mott states (CSM and PMI, respectively).
We can easily understand these relationships by comparing the schematic images shown in Fig. \ref{fig_SF}.
We comment on related studies on the system that deviates from half filling with an imbalanced population. 
The CSAF-like ordered state can appear, if the total filling factor of two of the three colors is integer.\cite{HH2004a,Rapp2011}

\subsection{Stability of staggered states}\label{sec_Mag2}
\begin{figure}[tb]
\begin{center}
\includegraphics[width=0.33\linewidth]{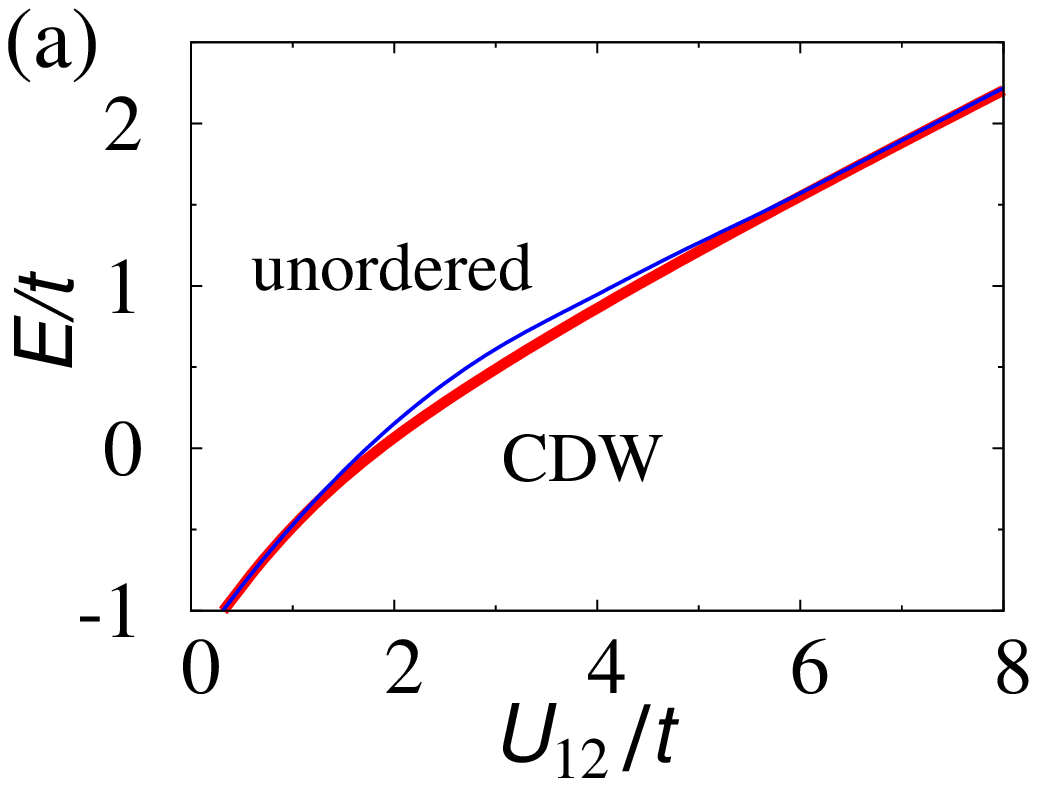}
\includegraphics[width=0.33\linewidth]{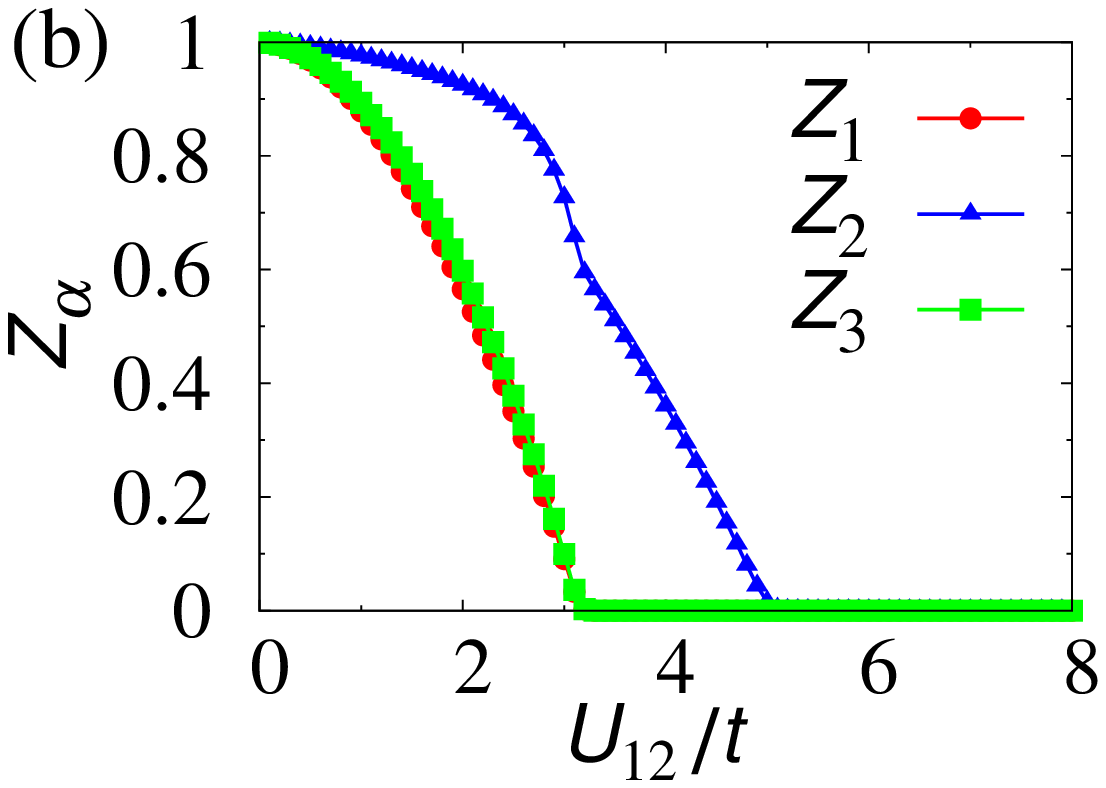}
\caption{(a) Energy $E/t$ for CDW (thick red line) and unordered states (thin blue line) and (b) renormalization factor $Z_\alpha$ for unordered states as functions of $U_{12}/t$ for $U_{23}/U_{12}=0.6$ and $U_{31}/U_{12}=2.0$. Calculations are for half filling at $T=0$.\protect\cite{Miyatake2010}
}
\label{fig_EvsU}
\end{center}
\end{figure}

Here, we discuss the stability of the staggered ordered states. 
For this purpose, we show the internal energy per site, $E\equiv \langle \hat{\cal H}\rangle/L$, calculated at $T=0$.
For comparison, we again consider the unordered states, the Mott and the FL states, obtained by using the same procedures as described in Sec. \ref{sec_Mott}. 
Figure \ref{fig_EvsU}(a) shows the energies of the CDW state and the unordered states for $U_{23}/U_{12}=0.7$ and $U_{31}/U_{12}=1.5$.
Figure \ref{fig_EvsU}(b) shows the renormalization factors $Z_\alpha$ as functions of $U_{12}/t$, which characterize the phase transitions between the unordered states: the FL, the CSM, and the PMI phases. 
From Fig. \ref{fig_EvsU}(a), we can see that the CDW is more stable than the unordered states, suggesting that the CDW is the ground state at $T=0$. 
As the temperature increases, the Mott states and the FL state become stable, which is discussed in Sec. \ref{subsec_Mag3}.

%---------------------------------------
%---------------------------------------

%
%---------------------------------------
\begin{figure}[tb]
\begin{center}
\includegraphics[height=3.0cm]{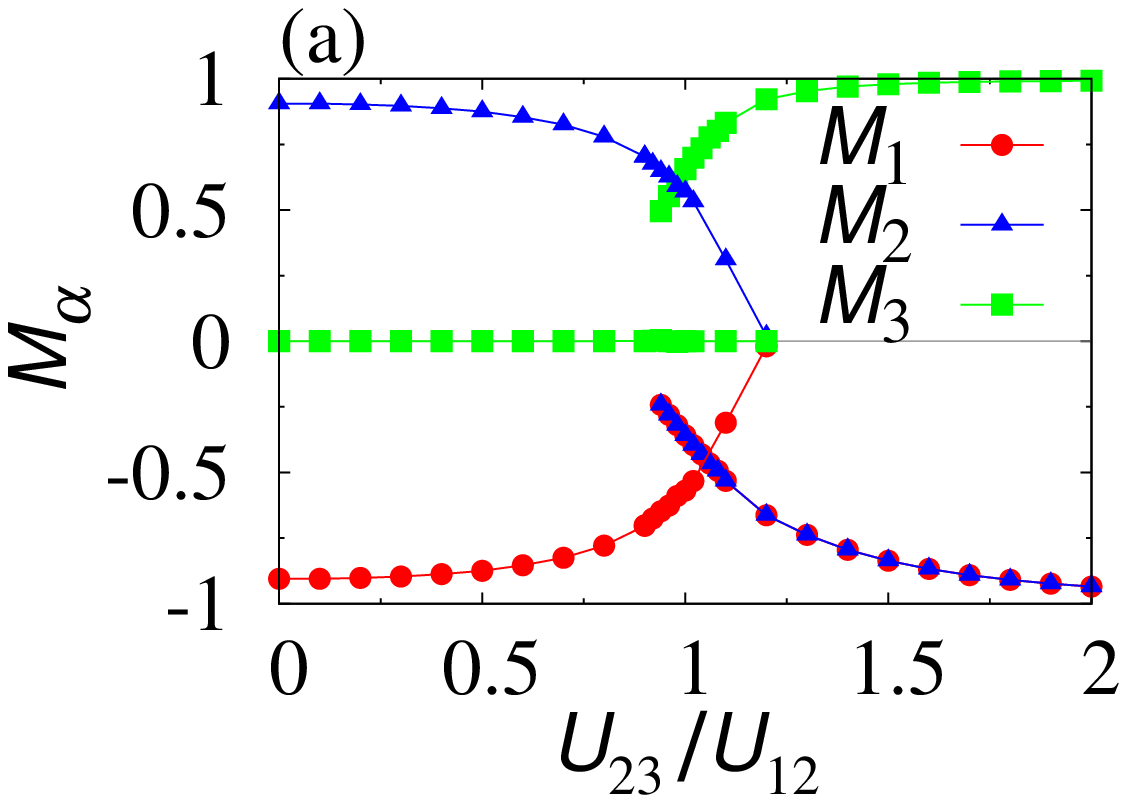}
\includegraphics[height=3.0cm]{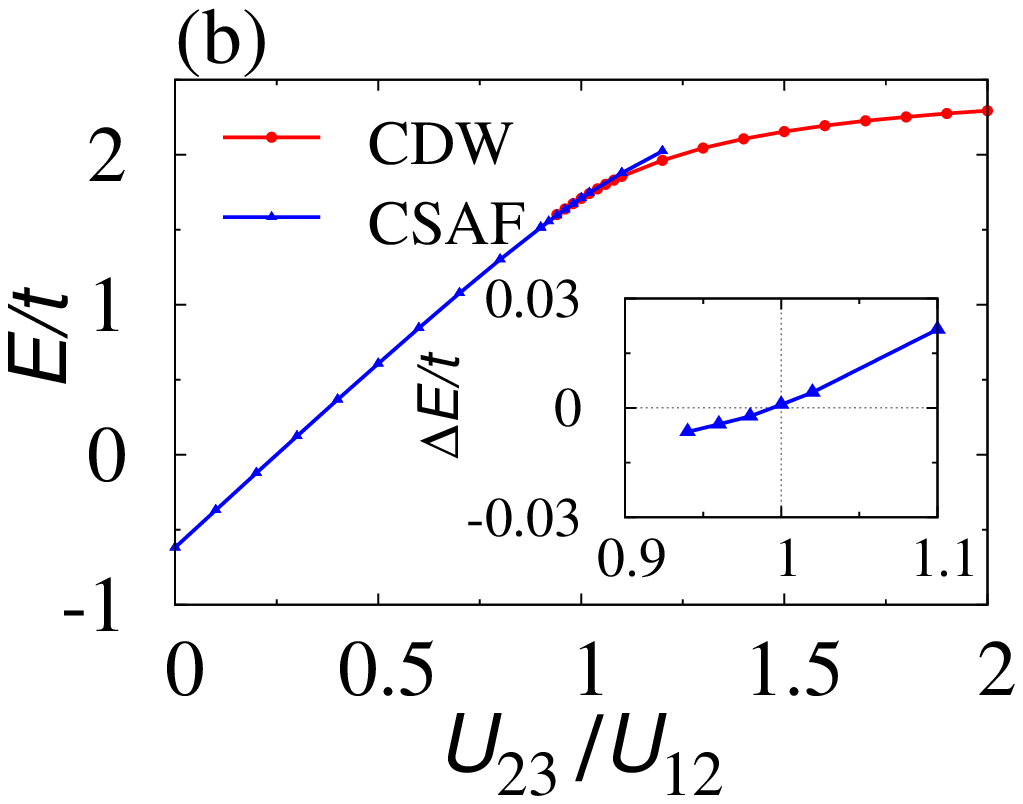}
\caption{(a) Order parameters and (b) energies for $U_{31}=U_{23}$ and $U_{12}/t=5$ as functions of $U_{23}/U_{12}$.\protect\cite{Miyatake2010} 
Inset: The energy difference $\Delta E\equiv E_{\rm CSAF}-E_{\rm CDW}$ between $E_{\rm CSAF}$ the energy of the CSAF and $E_{\rm CDW}$ that of the CDW.  
Calculations are for half filling at $T=0$.}
\label{fig_Deg}
\end{center}
\end{figure}
%---------------------------------------

Next, we focus on the stability of the ordered states at the SU(3) point ($U_{12}=U_{23}=U_{31}$). 
As mentioned in Sec. \ref{sec_Mott}, the Mott states are absent at this point. 
It is thus interesting to investigate what kind of ordered state becomes stable at this specific parameter point. 
We can expect that the CDW and the CSAF are two candidates for the ordered state at half filling.
We thus analyzed the half-filled system by varying $U_{23}/U_{12}$ for $U_{23}=U_{31}$ at $T=0$. In Fig. \ref{fig_Deg}, we show the order parameters $M_\alpha$ and the energy $E$ as functions of $U_{23}/U_{12}$. 
The CSAF appears when $U_{23}/U_{12}<1.20$, while the CDW appears when $U_{23}/U_{12}>0.94$.
Interestingly, we also find a hysteresis when $0.94<U_{23}/U_{12}<1.20$, implying a first-order quantum phase transition between the two ordered states. 
By comparing the energies of the two states shown in Fig. \ref{fig_Deg}(b), we can see that the energy of the CSAF and that of the CDW cross at around the SU(3) point.
These results indicate that the CSAF (CDW) is stable for $U_{23}/U_{12}<1$ ($U_{23}/U_{12}>1$) and that the two ordered states degenerate at the SU(3) point within our numerical accuracy. 
It should be noted that the CSAF and the CDW correspond to spontaneously symmetry-broken states induced by infinitesimal perturbations proportional to $\vec{a_{i}}\,^\dag {\bf t}_a \vec{a_{i}}$ and $\vec{a_{i}}\,^\dag {\bf t}_b \vec{a_{i}}$, respectively, where ${\bf t}_\ell$ are generators of the SU(3) group defined by 
$$
{\bf t}_{a}=\left(\begin{array}{ccc}
1&0&0\\
0&-1&0\\
0&0&0
\end{array}\right),
%\,\,{\rm and}\,\,\,
\,\,\,\,\,
{\bf t}_{b}=\frac{1}{\sqrt{3}}\left(\begin{array}{ccc}
1&0&0\\
0&1&0\\
0&0&-2
\end{array}\right),\,\,\,
{\rm and}\,\,\,
\vec{a}_{i}=\left(\begin{array}{c}
\hat{a}_{i1}\\
\hat{a}_{i2}\\
\hat{a}_{i3}
\end{array}\right).
$$
We thus conclude from our numerical results that the CSAF and the CDW are degenerate at the SU(3) point, which is supported in terms of the SU(3) symmetry.

\subsection{$U_{12}$-$T$ Phase diagram at half filling}\label{subsec_Mag3}

We review the phase transitions between the Mott states and the staggered ordered state driven by thermal fluctuations. 
Here, we present  results obtained with the self-energy functional approach (SFA),\cite{Potthoff03a,Potthoff03b} which is a reliable method for investigating the finite-temperature properties of an infinite-dimensional system.
This method allows us to accurately calculate various thermodynamic quantities  including the grand potential per site $\Omega$ and the entropy per site $S\equiv -\partial\Omega/\partial T$, where $\Omega=E-TS$.

\begin{figure}[tb]
\begin{center}
\includegraphics[scale=0.4]{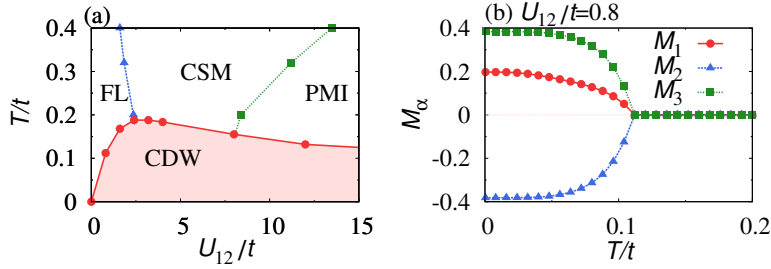}
\caption{(a) $U_{12}$-$T$ phase diagram and (b) temperature dependence of order parameter $M_{\alpha}$ for $U_{23}/U_{12}=0.6$ and $U_{31}/U_{12}=2$ at half filling obtained by SFA.\protect\cite{Suga2011}
The phase transition between the CDW and the FL is of the first order. The phase transitions between the CDW and the CSM and between the CDW and PMI are of the second order. The lines separating FL-CSM and CSM-PMI denote crossovers. 
}
\label{fig_UTphase_HF}
\end{center}
\end{figure}

Before proceeding to details, let us show the typical $U_{12}$-$T$ phase diagram [Fig. \ref{fig_UTphase_HF}(a)] determined for fixed $U_{23}/U_{12}=0.6$ and $U_{31}/U_{12}=2$. 
The ground state at $T=0$ is the CDW except when $U_{12}/t=0$, and the CDW phase extends at low temperatures. 
As shown in Fig. \ref{fig_UTphase_HF}(b), the order parameters $|M_\alpha|$ decrease as $T/t$ increases, suggesting the second-order phase transition from the CDW to the FL.
As discussed below, the phase transitions from the CDW to the CSM and the PMI occur as $T/t$ increases in the stronger $U_{12}/t$ region. 
At higher temperatures, we can see crossovers between the unordered states as $U_{12}/t$ changes.
These features obtained by the SFA agrees qualitatively with those mentioned in Sec. \ref{sec_Mott}.
We note that a similar phase diagram consisting of the CSAF, the CSM, and the FL is obtained also for, {\it e.g.}, $U_{12}=U_{23}$ and $U_{31}>U_{12}$.

%---------------------------------------
%---------------------------------------
%---------------------------------------
\begin{figure}[tb]
\begin{center}
\includegraphics[scale=0.45]{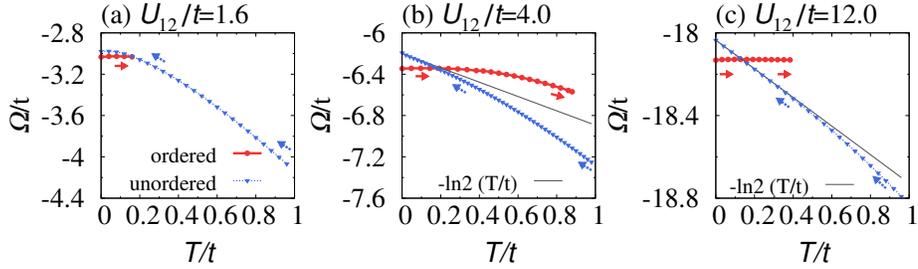}
\caption{Grand potential per site $\Omega$ as a function of temperature $T/t$ with fixed $U_{23}/U_{12}=0.6$ and $U_{31}/U_{12}=2$ for different $U_{12}/t=1.6$ (a), 4.0 (b) and 12.0 (c).\protect\cite{Suga2011} 
As shown by the arrows, the $\Omega$ curve for ordered (unordered) solutions are obtained with calculations starting from zero (high) temperature.
}
\label{fig_OmgvsT}
\end{center}
\end{figure}
%---------------------------------------
Let us move on to detailed calculations.
In Fig. \ref{fig_OmgvsT}, we show $\Omega/t$ as a function of $T/t$ for $U_{12}/t=1.6, 4.0$, and $12.0$ by fixing $U_{23}/U_{12}=2$ and $U_{31}/U_{12}=0.6$. Here, the red circles represent the results calculated with increasing $T/t$ from $0$, and the blue triangles are calculated with decreasing $T/t$ from 1. 
These two $\Omega/t$ curves correspond to the grand potentials for the CDW and the unordered states, respectively.
We checked that the order parameters $M_\alpha$ were finite for the calculations with increasing $T/t$ (not shown).
By comparing $\Omega/t$ curves for the CDW and the unordered states, we can analyze the stability of these states at finite temperatures. 
At $T=0$, the grand potential is equivalent to the internal energy. 
Thus, we can see that the CDW state is stable there as discussed in Sec. \ref{sec_Mag2}.

For $U_{12}/t=1.6$ [Fig. \ref{fig_OmgvsT}(a)], $\Omega/t$ for the CDW has little dependence on $T/t$.
This feature suggests that the CDW has a spectral gap.\cite{Miyatake2010}
In contrast,  $\Omega/t$ for the unordered states strongly depends on $T/t$, which characterizes the FL, the CSM, or the PMI.
We can see that $\Omega/t$ for the unordered states at $U_{12}/t=1.6$ obeys $\Omega/t \propto (T/t)^2$. 
This temperature dependence yields the expression $S\propto T/t$, which characterizes the FL.\cite{Georges1996,Potthoff03a,Potthoff03b}
As $T/t$ increases, the $\Omega/t$ curves for the CDW and the FL connect smoothly with each other at $T/t\sim0.18$, at which point the CDW disappears.
This suggests that the second-order phase transition between the FL and the CDW occurs at this temperature. 
The second-order phase transition is confirmed by the temperature dependence of the order parameter in a smaller $U_{12}/t$ region shown in Fig. \ref{fig_UTphase_HF}(b).

For $U_{12}/t=4.0$ [Fig. \ref{fig_OmgvsT}(b)], the unordered state exhibits $\Omega/t \sim -a_{}(T/t)^2 -(\ln2) T/t$ with a positive constant $a$.
The entropy per site is thus written as $S \sim 2aT/t + \ln2$, characterizing the CSM as follows.
The first term stems from the contribution of the itinerant atoms. 
The second term is due to the localized atoms of two colors. 
Namely, the entropy of $\ln 2$ results from the free local $S=1/2$-pseudospins as mentioned in Sec. \ref{subsec_discuss_Mott}.
The two $\Omega/t$ curves cross at $T/t \sim 0.18$, indicating the first-order phase transition between the CDW and the CSM.

For $U_{12}/t=12.0$ [Fig. \ref{fig_OmgvsT}(c)], the $\Omega/t$ curves for the CDW and the unordered states cross at $T/t \sim 0.13$, indicating that the first-order phase transition occurs at this point. 
For $T/t<0.4$, the unordered state shows $\Omega/t \sim -(\ln2) T/t$, yielding $S \sim \ln 2$. 
This result means that the unordered state is the PMI, where each site is occupied by either the paired atoms or the third-color atom. 
This two-fold degeneracy at each site yields the entropy of $\ln 2$ as discussed in Sec. \ref{subsec_discuss_Mott}.
As $T/t$ is increased, the $\Omega/t$ curve for the unordered state gradually changes to $\Omega/t \sim -a_{}(T/t)^2 -(\ln2) T/t$ with $a>0$ at $T/t\sim 0.4$, indicating that the crossover from the PMI to the CSM occurs due to thermal fluctuations.
Note that a state with a higher entropy becomes stable at higher temperatures because of the term  $-ST$ in $\Omega = E - ST$.

%---------------------------------------
\begin{figure}[tb]
\begin{center}
\includegraphics[scale=0.42]{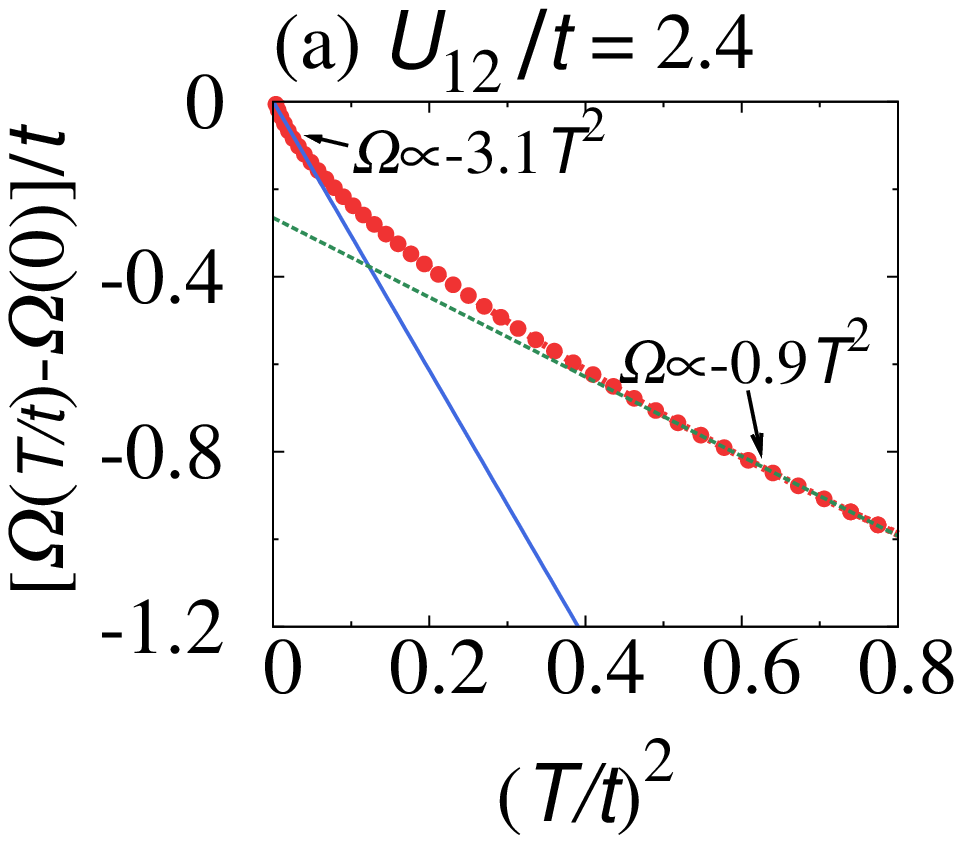}
\includegraphics[scale=0.42]{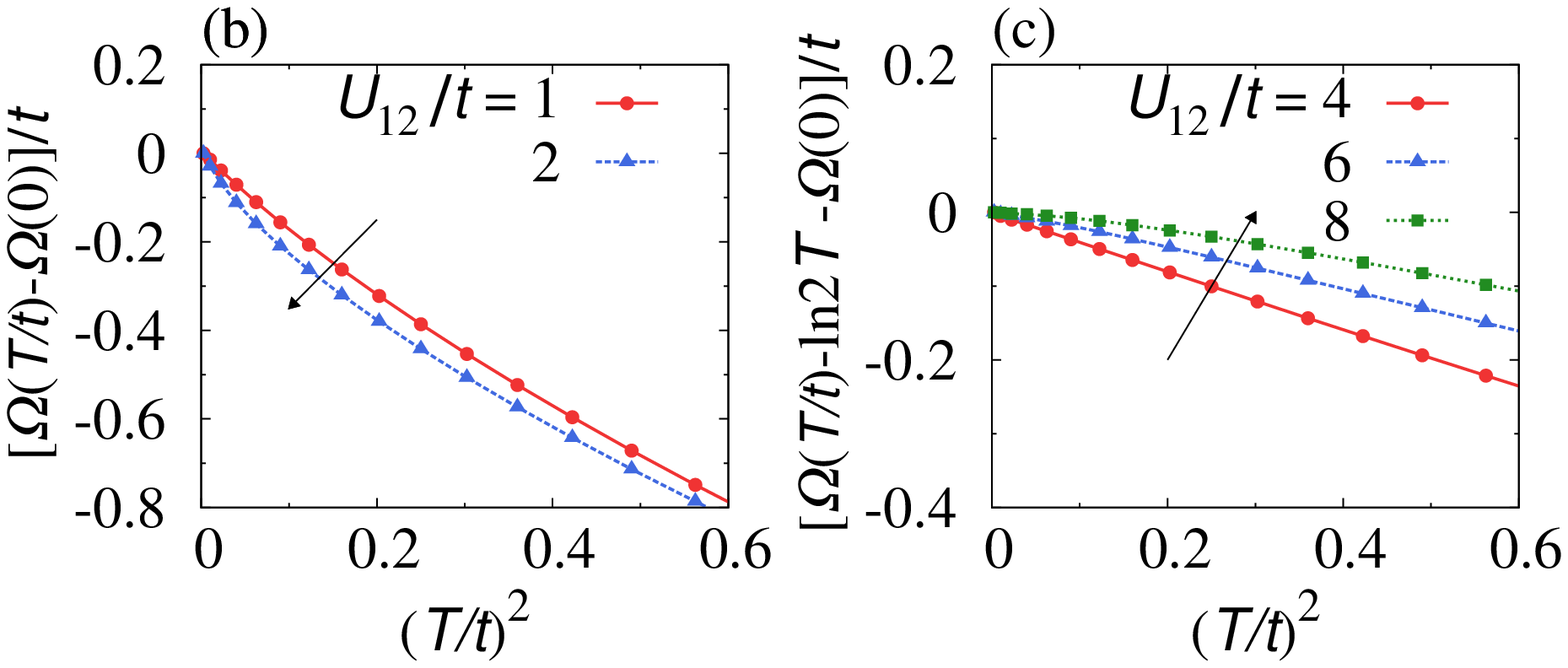}
\caption{(a) Grand potential per site for unordered states as a function of squared temperature $(T/t)^2$ for $U_{12}/t=2.4$ with $U_{23}/U_{12}=0.6$ and $U_{31}/U_{12}=2$, where grand potential is measured from that at zero temperature defined as $\tilde{\Omega}(T/t)\equiv \Omega(T/t)-\Omega(0)$.
Two thin lines represent $\tilde{\Omega}/t \propto (T/t)^2$ with different coefficients. (b) and (c) $\tilde{\Omega}$ as a function of $(T/t)^2$ for different $U_{12}/t$ values with $U_{23}/U_{12}=0.6$ and $U_{31}/U_{12}=2$.
The contribution $-(\ln 2)T$ from the localized atoms is subtracted from $\tilde{\Omega}$ in the panel (c).}
\label{fig_NFL_HF}
\end{center}
\end{figure}
%---------------------------------------

Below, we discuss another crossover between the FL and the CSM in a weakly interacting region. 
Figure \ref{fig_NFL_HF}(a) displays $\tilde{\Omega}(T/t)$ $[\equiv \Omega(T/t)-\Omega(0)]$ for unordered states as a function of $(T/t)^2$ at $U_{12}/t=2.4$. 
We find that $\tilde{\Omega}/t$ depends linearly on $(T/t)^2$ for both small and large $(T/t)^2$. 
On the other hand, the slope of the linear behavior in the lower temperature region is nearly three times larger than that in the higher temperature region. 
This factor of three is attributed to the difference in the number of  colors of itinerant atoms. 
Thus, we can conclude that the crossover with increasing $T/t$ is from the FL to the CSM.

Here, we discuss non-FL  behavior in the CSM. 
The temperature dependence of $\tilde{\Omega}/t \propto (T/t)^2$ in the CSM is caused by the itinerant atoms. 
In the small $U_{12}/t$ region shown in Fig. \ref{fig_NFL_HF}(b), the magnitude of its coefficient, which is proportional to the effective mass of the itinerant atom, increases as $U_{12}/t$ increases. 
This is typical FL behavior. 
However, in the strong $U_{12}/t$ region shown in Fig. \ref{fig_NFL_HF}(c), the magnitude of the coefficient decreases as $U_{12}/t$ increases. 
This curious $U_{12}/t$ dependence indicates that itinerant atoms in the CSM do not behave in the same way as the usual FL. 
As explained in Sec. \ref{subsec_app_Mott}, itinerant atoms in the CSM are effectively described by the Falikov-Kimball model. 
In this model, itinerant fermions have finite lifetime due to the scattering by localized fermions, and the usual FL behavior breaks down.

The analyses so far have been restricted to the homogeneous system. In the PMI and the CDW, the pair formation mechanism involves a potential instability towards the phase separation shown in Fig. \ref{fig_SF}(i). 
The competition between the PMI and CDW, and the phase separation is left as an open issue.

%%%%%%%%%%%%%%%%%%%%%%%%%%%%%%%%%%%%%%%%%%%%%%
\section{Superfluidity}\label{sec_SF}
%%%%%%%%%%%%%%%%%%%%%%%%%%%%%%%%%%%%%%%%%%%%%%
In this section, we discuss the superfluidity of a {\it repulsively} interacting three-component system.
As presented in Sec. \ref{sec_Mott}, the characteristic Mott transition occurs, which is accompanied by effective pair formation at half filling. 
In particular, when one of the three interactions is weak, for instance, $U_{12}\ll U_{23}=U_{31}$, the FL phase appears close to the  PMI phase.
This FL region is strongly influenced by fluctuations resulting from the pair formation. 
When we take account of the appearance of the staggered ordered state at half filling, the ground state can be the CDW or the CSAF as discussed in Sec. \ref{subsec_discuss_Mott} and Sec. \ref{subsec_app_Mag}. The FL region with strong fluctuations of the pair formation thus appears not at half filling but close to half filling in the CDW/PMI parameters.

These findings and consideration motivated us to investigate the exotic superfluid state induced by the repulsive interactions in the above specific parameter region.
We have indeed found stable superfluid states in $U_{12} \ll U_{23}=U_{31}$ close to half filling.\cite{Inaba2012,Suga2012}
We have finally obtained finite-temperature phase diagrams.

\subsection{Origin of superfluidity}\label{subsec_eff}
\begin{figure}[tb]
\begin{center}
\includegraphics[scale=0.45]{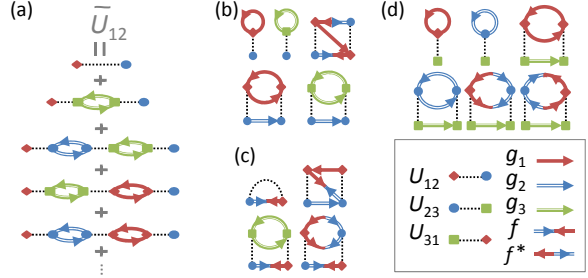}
\caption{(a) Diagrammatic representation of the effective interaction between color-1 and 2 atoms. Diagrams for the (b) normal and (c) anomalous self-energies of color-1 and 2 atoms, respectively, and (d) the normal self-energies of color-3 atoms.
}
\label{fig_rpa}
\end{center}
\end{figure}

We first analyze effective interactions renormalized by  many-body effects using a random phase approximation. 
This is a standard method for investigating the origin of  superfluidity.
We consider the effective interaction between color-1 and 2 atoms, $\tilde{U}_{12}$.
Figure \ref{fig_rpa}(a) is a diagrammatic representation of $\tilde{U}_{12}$. 
The collected bubble-type diagrams correspond to the density fluctuations of the atoms, and thus the random phase approximation captures the renormalization effects of such types of fluctuations.
% that we collected based on the random phase approximation.
The contribution of these diagrams is summarized in the following expression,  
%****************************************************************
\begin{eqnarray}
\tilde{U}_{12}({\bf x})&=&\Big[U_{12}-U_{31}\chi_3({\bf x})U_{23}\Big]
\bigg[1 -U_{12}^2\chi_1({\bf x})\chi_2({\bf x}) 
-U_{31}^2\chi_1({\bf x})\chi_3({\bf x})  \nonumber \\
&&-U_{23}^2\chi_2({\bf x})\chi_3({\bf x}) 
+2U_{12}U_{23}U_{31}\chi_1({\bf x})\chi_2({\bf x})\chi_3({\bf x})\bigg]^{-1},
\end{eqnarray}
%****************************************************************
%\begin{widetext}
%\begin{eqnarray}
%\tilde{U}_{12}({\bf x})=\frac{U_{12}-U_{31}\chi_3({\bf x})U_{23}}
%{1 -U_{12}^2\chi_1({\bf x})\chi_2({\bf x})  -U_{31}^2\chi_1({\bf x})\chi_3({\bf x})  -U_{23}^2\chi_2({\bf x})\chi_3({\bf x})
%+2U_{12}U_{23}U_{31}\chi_1({\bf x})\chi_2({\bf x})\chi_3({\bf x})},
%\end{eqnarray}
%\end{widetext}
%****************************************************************
where ${\bf x}=(i\nu_l,{\bm k})$, $\nu_l$ is the bosonic Matsubara frequency, and ${\bm k}$ is the wave vector. 
A bubble diagram corresponding to density fluctuations, $\chi_\alpha({\bf x})$, is given by $-\sum_{\bf y}g_\alpha({\bf y})g_\alpha({\bf y+x})$, where ${\bf y}=(i\omega_n,{\bm q})$ with $\omega_n$ and ${\bm q}$ being the fermionic Matsubara frequency and the wave vector, respectively, and the bare Green's function of color-$\alpha$ atoms is denoted by $g_\alpha({\bf y})=(i\omega_n+\mu_\alpha-\varepsilon_{\bm q})^{-1}$. 
The other effective interactions can be evaluated in the same way. 
The conditions $U_{12}\equiv U_{12}$ and $U_{31}=U_{23}\equiv U_{23}$ yield $\chi_1=\chi_2 \equiv \chi$ and $\chi_3 \equiv \chi'$, and $\mu_1=\mu_2\equiv \mu$ and $\mu_3\equiv \mu'$. 
Thus, the effective interactions $\tilde{U}({\bf x})\left[\equiv\tilde{U}_{12}({\bf x})\right]$ and $\tilde{U}'({\bf x})\left[\equiv\tilde{U}_{23}({\bf x})=\tilde{U}_{31}({\bf x})\right]$ are reduced to
%****************************************************************
\begin{eqnarray}
\tilde{U}({\bf x})&=&\frac{U_{12}-U_{23}^2\chi'({\bf x})}
{[1-U_{12}\chi({\bf x})][1+U_{12}\chi({\bf x})-2U_{23}^2\chi({\bf x})\chi'({\bf x})]}, 
\label{U}\\
\tilde{U}'({\bf x})&=&\frac{U_{23}}
{1+U_{12}\chi({\bf x})-2U_{23}^2\chi({\bf x})\chi'({\bf x})}. 
\label{U_{23}}
\end{eqnarray}
%\begin{eqnarray}
%\tilde{U}({\bf x})&=&\frac{-U_{23}^2\chi'({\bf x})}
%{1-2U_{23}^2\chi({\bf x})\chi'({\bf x})}\nonumber,\
%\tilde{U_{23}}({\bf x})&=&\frac{U_{23}}
%{1-2U_{23}^2\chi({\bf x})\chi'({\bf x})}\nonumber,
%\end{eqnarray}
%****************************************************************
The condition $U_{12} \ll U_{23} < t$ further simplifies $\tilde{U}({\bf x})$ as $\sim U_{12} -U_{23}^2\chi'({\bf x})$. 
We should note that the second term is attractive. 
Therefore, the effective interaction $\tilde{U}({\bf x})$ can become attractive if the second term overcomes the bare repulsion $U_{12}$. 
In contrast, for $U_{12}=U_{23}< t$, the effective interaction is repulsive, $\tilde{U}({\bf x})=\tilde{U}'({\bf x}) \sim U_{12}$. 
%****************************************************************
%****************************************************************

The simple analysis suggests that the effective attractive interaction between color-1 and 2 atoms is induced by density fluctuations of the color-3 atoms when the anisotropy is strong ($U_{12}/U_{23} \ll 1$). 
We emphasize again that the condition $U_{12}/U_{23} \ll 1$ with $U_{23}=U_{31}$ is essentially equivalent to the condition that must be met for the appearance of the direct FL-PMI transition in the strongly correlated region [see Fig. \ref{fig_RU_phase}(b)]. 
This indicates that the $s$-wave superfluid can be induced by the attractive $\tilde{U}({\bf x})$ close to the PMI transition point at low temperatures.

%
%****************************************************************
\begin{figure}[tb]
\begin{center}
\includegraphics[scale=0.45]{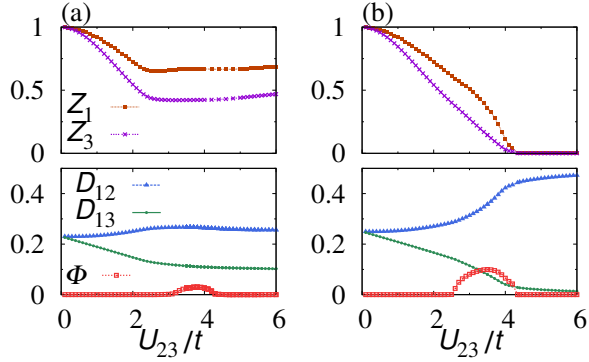}
\caption{
Renormalization factors $Z_\alpha$, superfluid order parameter $\Phi$, and double occupancies $D_{\alpha\beta}$ calculated with the modified IPT as functions of $U_{23}/t$ at $U_{12}/U_{23}=0.1$ and $T/t=0.03$ for fillings (a) $N=1.44$ and (b) $N=1.5$.\protect\cite{Inaba2012} 
}
\label{fig_udep}
\end{center}
\end{figure}
%****************************************************************

\subsection{Appearance of superfluid state}
We further discuss the appearance of the superfluid state based on numerical simulations.
We employed a modified version of the iterative perturbation theory (IPT),\cite{Garg2005,Inaba2012} which is a combination of the DMFT and the diagrammatic perturbation method. 
We iteratively calculated the effective impurity problem with a perturbative summation up to the second-order Feynman diagrams.
Figure \ref{fig_rpa}(b), (c), and (d) represent a collection of  diagrams that take account of the competition between the bare repulsion $U_{12}$ and the effective attraction $-U_{23}^2\chi(\omega)$.
We used this method to calculate the superfluid order parameter $\Phi=\langle \hat{a}_{i1}\hat{a}_{i2} \rangle$ in addition to $Z_\alpha$ and $D_{\alpha \beta}$.

Figure \ref{fig_udep}(a) represents the $U_{23}/t$ dependence of $\Phi$, $Z_\alpha$, and $D_{\alpha\beta}$ at  $U_{12}/U_{23}=0.1$ and $T/t=0.03$ for $N=1.44$.
The superfluid order parameter $\Phi$ smoothly increases from zero at $U_{23}/t\sim 3$ and then smoothly disappears at $U_{23}/t\sim 4.2$.
This indicates that the superfluid state appears in the region $3<U_{23}/t<4.2$, and the transition between the FL and the superfluid is continuous.
Away from half filling, $Z_\alpha$ is always finite because the PMI does not appear due to the non-integer effective filling $n_{\rm pair}+n_3$.\cite{Inaba2010b,Gorelik} 
We note that the superfluid phase $3<U_{23}/t<4.2$ corresponds to the doped region close to the FL-PMI transition point.

For comparison, we present a result at half filling, $N=3/2$, without considering the possibility of staggered orders.
Figure \ref{fig_udep}(b) shows $\Phi$, $Z_\alpha$, and $D_{\alpha\beta}$ for the same parameters as above except for $N$.
From the order parameter $\Phi$, we can see that the superfluid state appears in the region $2.5<U_{23}/t<4.2$.
The $Z_\alpha$ curves also show that the PMI state appears for $U_{23}/t>4.2$, suggesting that the transition from the superfluid to the PMI state occurs at $U_{23}/t=4.2$.
When $U_{23}/t$ approaches the PMI transition point from below, $\Phi$ reaches zero with a tiny jump at the PMI transition point $U_{12}/t\sim 4.2$. 
Thus, the superfluid-PMI transition is  discontinuous unlike the FL-superfluid transition.

Let us summarize the mechanism of the superfluid transition.
Due to the large anisotropy $U_{12}/U_{23}$, the color-1 and 2 atoms tend to keep away from the color-3 atoms to avoid the strong $U_{23}/t$.
As mentioned in Sec. \ref{subsec_eff}, the density fluctuations of the third-color atoms mediate the effective attractive interaction that causes the characteristic superfluid.
On the other hand, the pair-formation mechanism can also induce instability toward a phase separation in the large $U_{23}/t$ region. 
In the next subsection, we discuss the validity of the above results by analyzing the stability of the superfluid state.

\subsection{Stability of superfluid state}
\begin{figure}[tb]
\begin{center}
\includegraphics[scale=0.45]{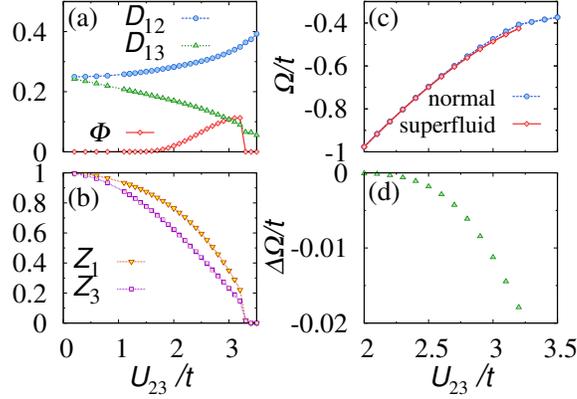}
\caption{(a) and (b) $\Phi$, $Z_\alpha$ and $D_{\alpha\beta}$ obtained with the self-energy functional approach using the same parameter region as that used in Fig. \ref{fig_udep}(b).\protect\cite{Inaba2012} 
(c) Comparison of grand potential per site $\Omega$ between the superfluid state and the normal fluid state at zero temperature.\protect\cite{Inaba2012} 
(d) Difference of grand potentials defined by $\Delta\Omega=\Omega_{\rm superfluid}-\Omega_{\rm normal}$. 
}
\label{fig_SFA_SF}
\end{center}
\end{figure}

We first checked the validity of the modified IPT calculations that take into account certain Feynman diagrams only. 
For this purpose, we performed the unperturbative calculations with the SFA at half filling, $N=3/2$.  
We comment that it is difficult to systematically analyze the superfluid state for filling away from $N=3/2$ using the SFA. 
This is because the SFA is based on a variational method. 
Particle-hole symmetry at $N=3/2$ allows us to reduce the number of variational parameters, while the convergency of the method becomes worse when the number of variational parameters increases for $N\not=3/2$.
Figure \ref{fig_SFA_SF}(a) and (b) represent $D_{\alpha\beta}$, $\Phi$ and $Z_\alpha$ calculated by the SFA for the same parameters as  in Fig. \ref{fig_udep}(b). 
By comparing these results, we find that the SFA and the IPT are in good qualitative agreement. 
We thus stress that the IPT diagrams in Fig. \ref{fig_rpa}(b) capture the essential properties of the superfluid transition.

Let us move on to the stability of the superfluid state, which we confirmed by examining the grand potential per site, $\Omega$. 
In the same way as discussed in Sec. \ref{subsec_Mag3}, we calculated $\Omega$ for the normal states by neglecting the possible ordered states. 
Figure \ref{fig_SFA_SF}(c) displays $\Omega$ for the superfluid state and that for the normal states, the FL, and the PMI. 
Figure \ref{fig_SFA_SF}(d) displays their difference defined by $\Delta\Omega=\Omega_{\rm superfluid}-\Omega_{\rm normal}$. 
We can see that the superfluid state has a smaller grand potential, $\Omega$, than the normal states. 
%\add{This result is reasonable in terms of the formation of a superfluid gap of color-1 and 2 atoms around the Fermi energy (see Fig. \ref{fig_dos}(b) below). }
This indicates the stability of the superfluid state.

%****************************************************************
%****************************************************************

%
%****************************************************************
\begin{figure}[tb]
\begin{center}
\includegraphics[scale=0.45]{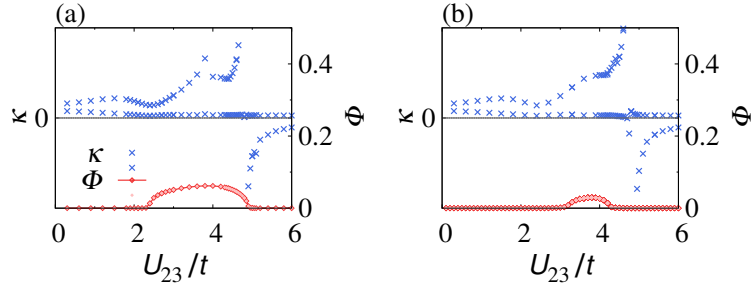}
\caption{Eigenvalues ($\kappa$) of $\kappa_{ab}$ at (a) $T/t=0.02$ and (b) 0.03  for $U_{12}/U_{23}=0.1$ and $N=1.44$.\protect\cite{Inaba2012} 
Negative values denote the instability of the homogeneous phases. 
The order parameter $\Phi$ is also plotted, for comparison. 
}
\label{fig_kapper}
\end{center}
\end{figure}
%****************************************************************
We further discuss the stability of the superfluid phase against the phase separation.
Recent theoretical studies on the attractively interacting three-component atoms pointed out the instability toward a phase separation.\cite{Priv,Titv} 
It was shown that, in a strongly attractive region, the trionic state becomes unstable and  phase separation occurs; one phase consists of two-color atoms that form Cooper pairs, while the other consists of the third-color atoms. 
In the present repulsive case, the same type of instability can occur as a result of the pair formation. 
We thus investigated the possibility of the phase separation between paired color-1 and 2 atoms, and unpaired color-3 atoms.
We calculated the $2\times2$ compressibility matrix 
%****************************************************************
%\begin{eqnarray}
$\kappa_{ab}\equiv \frac{\partial N_a}{\partial\mu_b},$
%\end{eqnarray}
%****************************************************************
where $N_a=n_1+n_2, n_3$, and diagonalized it. Negative eigenvalue of the compressibility matrix indicates instability toward the phase separation.
Figure \ref{fig_kapper} shows the eigenvalues $\kappa_{}$ at $T/t=0.02$ and 0.03 for $U_{12}/U_{23}=0.1$ and $N=1.44$. 
We find that, for both $T/t=0.02$ and $0.03$, one of the eigenvalues exhibits an anomaly at $U_{23}/t \sim 4.8$ and is negative when $U_{23}/t \gtrsim 4.8$. 
Accordingly, the phase separation occurs in $U_{23}/t \gtrsim 4.8$. 
In the phase made up of paired color-1 and 2 atoms, the superfluidity disappears in contrast to the attractive case, because the effective attractive interaction is no longer induced by unpaired color-3 atoms that are separated in space. 
The stable homogeneous superfluid and FL phases are found when $U_{23}/t \lesssim 4.8$.

\subsection{\it Dynamical properties of superfluid state}
%****************************************************************
\begin{figure}[tb]
\begin{center}
\includegraphics[scale=0.5]{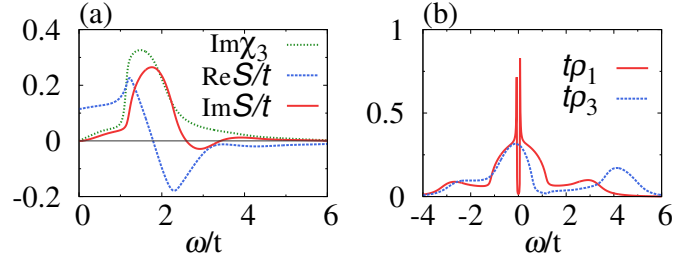}
\caption{
(a) Anomalous self-energy ${\rm Re}S(\omega_+)$, ${\rm Im}S(\omega_+)$, and imaginary part of the polarization function ${\rm Im}\chi_3(\omega_+)$.\protect\cite{Inaba2012} 
(b) Single-particle excitation spectra $\rho_\alpha(\omega)$.\protect\cite{Inaba2012} 
These quantities are calculated with the modified IPT.  Parameters are $U_{23}/t=3$, $U_{12}/U_{23}=0.1$, $N=1.44$, and $T/t=0.01$.
}
\label{fig_dos}
\end{center}
\end{figure}
%****************************************************************

Here, we present the dynamical properties of the superfluid state, which provides a further insight into the origin of superfluidity.
As discussed using Eq. (\ref{U}), the density fluctuations of color-3 atoms play a key role in the appearance of a superfluid. 
Thus, it is very informative to calculate ${\rm Im}\chi_3(\omega)$ and the anomalous self-energy $S(\omega)$. 
In Fig. \ref{fig_dos}(a), we show ${\rm Im}\chi_3(\omega_+)$, ${\rm Re}S(\omega_+)$, and ${\rm Im}S(\omega_+)$ with $\omega_+\equiv\omega+i0^+$ for $U_{23}/t=3$, $U_{12}/U_{23}=0.1$, $N=1.44$, and $T/t=0.01$. 
The spectra ${\rm Im}\chi_3(\omega)$ have a peak around $\omega_p/t\sim1.6$, indicating that the scattering caused by the density fluctuations of color-3 atoms occurs prominently around $\omega\sim\omega_p$. 
The anomalous self-energy $S(\omega)$ exhibits characteristic behavior at $\omega/t\sim1.8$ $(\gtrsim\omega_p/t)$; 
The real (imaginary) part  ${\rm Re}S(\omega_+)$ [${\rm Im}S(\omega_+)$] exhibits resonant behavior (peak structure). 
On the basis of the perturbative approach, ${\rm Re}S(\omega_+)$ is understood as a measure of the effective interaction.\cite{Scalapino} 
From the sign of ${\rm Re}S(\omega_+)$, we can thus see that the effective interaction between color-1 and 2 atoms is attractive for $\omega/t<1.8$.
The peak of ${\rm Re}S(\omega_+)$ appears at around a smaller frequency than $\omega_p/t$, and ${\rm Re}S(\omega_+)$ remains negative for a large $\omega/t$. 
These features are attributed to the bare repulsion $U_{12}$ between color-1 and 2 atoms.\cite{Scalapino} 
These dynamical properties are qualitatively the same as those of the strong-coupling superconductor caused by the electron-phonon interaction.\cite{Scalapino}

In Fig. \ref{fig_dos}(b), we show the single-particle excitation spectra defined as $\rho_\alpha(\omega)=-(1/\pi){\rm Im}G_{\rm \alpha}(\omega_+)$. 
The $s$-wave superfluid gap is seen around the Fermi energy ($\omega=0$) in $\rho_1(\omega)[=\rho_2(\omega)]$, while the renormalized peak is seen in $\rho_3(\omega)$ owing to the FL property. 
This $s$-wave spectral gap is consistent with the dynamical properties shown in Fig. \ref{fig_dos}(a).
We find that the incoherent Hubbard band structures appear in both  $\rho_1(\omega)$ and $\rho_3(\omega)$, which stem from strong correlation effects. 
Actually, the peak position of ${\rm Im}S(\omega_+)$ is inside this incoherent Hubbard-band region of $\rho_1(\omega)$, $|\omega/t|\gtrsim 1.5$. 
On the basis of the results described so far, we can stress that the effective attractive interaction between color-1 and 2 atoms is mediated by the density fluctuations of color-3 atoms.

\subsection{Finite-temperature phase diagrams close to half filling}
%%%%%%%%%%%%%%%%%%%%%%%%%%%%%%%%%%%%%%%%%%%%%%
%****************************************************************
\begin{figure}[tb]
\begin{center}
\includegraphics[scale=0.45]{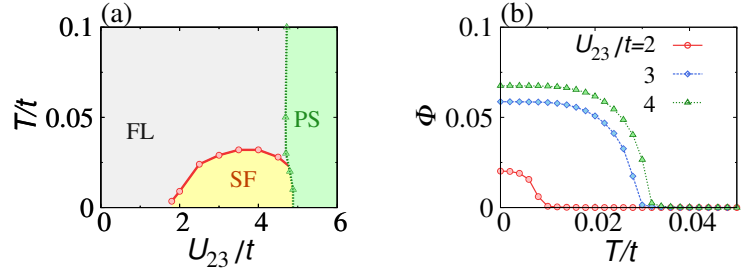}
\caption{
(a) $U_{23}$-$T$ phase diagrams for $U_{12}/U_{23}=0.1$ at $N=1.44$ obtained with the modified IPT.\protect\cite{Inaba2012} 
SF: superfluid; PS: phase separation.
(b) $\Phi$ as a function of $T/t$ for $U_{12}/U_{23}=0.1$ at $N=1.44$.\protect\cite{Inaba2012} 
}
\label{fig_phase}
\end{center}
\end{figure}
%****************************************************************
We finally discuss the finite-temperature phase diagrams obtained by systematically changing $T$ and $U_{23}$.
Figure \ref{fig_phase}(a) presents the $U_{23}$-$T$ phase diagram for $U_{12}/U_{23}=0.1$ at $N=1.44$ calculated with the modified IPT.
The superfluid phase appears in the moderately interacting region around $U_{23}/t \sim 4$, where the interaction and the bandwidth $4t$  are  comparable. 
In the strongly interacting region, the homogeneous phase becomes unstable, and  phase separation occurs. 
In Fig. \ref{fig_phase}(b), we show the temperature dependence of $\Phi$ for $U_{12}/U_{23}=0.1$ at $N=1.44$. 
We find that $\Phi \propto \sqrt{T-T_c}$ with $T_c$ being the superfluid transition temperature. 
This behavior is typical of weak- and strong-coupling phonon-mediated superconductors.\cite{Scalapino}

%****************************************************************
\begin{figure}[tb]
\begin{center}
\includegraphics[scale=0.45]{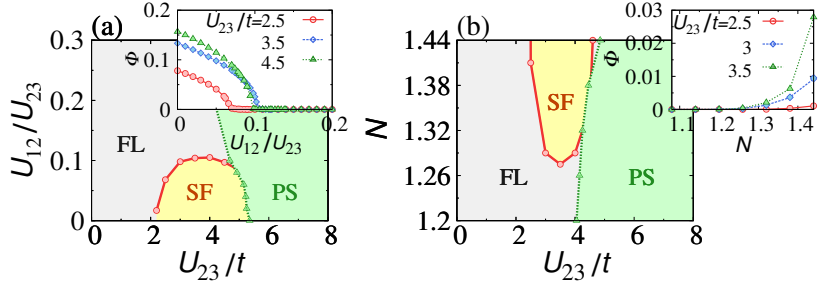}
\caption{
Phase diagrams obtained with the modified IPT at $T/t=0.03$.\protect\cite{Inaba2012} 
(a) $U_{23}$-$U_{12}$ phase diagram for $N=1.44$. The inset shows the $U_{12}/U_{23}$-dependence of the order parameter $\Phi$. 
(b) $U_{23}$-$N$ phase diagrams for $U_{12}/U_{23}=0.1$. The inset shows the $N$-dependence of the order parameter $\Phi$. 
}
\label{fig_phase_RU_NU}
\end{center}
\end{figure}
%****************************************************************
We also present the $U_{12}/U_{23}$- and $N$-dependent phase diagrams (Fig. \ref{fig_phase_RU_NU}). 
As shown in the inset of Fig. \ref{fig_phase_RU_NU}(a), $\Phi$ decreases and then vanishes as $U_{12}/U_{23}$ increases.
This is because the effective attractive interaction cannot overcome the bare repulsion $U_{12}$ in the weak anisotropy (large $U_{12}/U_{23}$) region as discussed based on Eq. (\ref{U}). 
We find that the superfluid appears for $U_{12}/U_{23} \lesssim0.11$ as shown in the main panel of Fig. \ref{fig_phase_RU_NU}(a), indicating that the strong anisotropy is necessary for the appearance of the superfluid. 
The inset of Fig. \ref{fig_phase_RU_NU}(b) shows that, as $N$ deviates from 1.5, $\Phi$ decreases monotonically due to the suppression of the density fluctuations of the color-3 atoms. 
We can see that the superfluid survives down to $N\sim 1.27$.

%%%%%%%%%%%%%%%%%%%%%%%%%%%%%%%%%%%%%%%%%%%%%%
\section{Summary and Outlook}\label{sec_Summary}
%%%%%%%%%%%%%%%%%%%%%%%%%%%%%%%%%%%%%%%%%%%%%%
In this review, we have discussed the novel ordered states and  Mott states that are unique to repulsively interacting three-component fermionic atoms in optical lattices. 
We first showed that the characteristic Mott transitions occur at half filling (3/2 atoms per site), when the interactions are anisotropic (color-dependent). 
We have found that  quantum phase transitions occur among the Fermi liquid (FL), the color-selective Mott state (CSM), and the paired Mott insulator (PMI). 
We have also found that two types of staggered ordered states, the color-density wave (CDW) and the color-selective antiferromagnet (CSAF), appear depending on the anisotropy of the interactions at half filling. 
These two staggered ordered states are degenerate for the isotropic SU(3) interactions within our numerical accuracy. 
We have then pointed out that the superfluid appears in the vicinity of the PMI transition point close to half filling. 
We have revealed that two-color atoms with the weakest repulsion form Cooper pairs, which are mediated by the density fluctuations of the third-color atoms.  
This Cooper-pairing mechanism is essentially similar to that of  conventional phonon-mediated superconductivity, although the superfluid phase appears close to the Mott insulating (PMI) phase. 
Knowledge of this novel superfluid would provide an insight into the Cooper pairing mechanism of other repulsively interacting systems in, for instance, strongly correlated electrons.

In the present study, we investigated infinite-dimensional bipartite Bethe lattice systems by using several numerical approaches based on the dynamical mean-field theory (DMFT). The DMFT calculations capture local correlation effects. 
We can thus expect the results obtained here regarding the Mott transition to be applicable to other bipartite systems.
Recent progress in creating optical lattices enables us to design the lattice geometry of low-dimensional systems. 
For the two-dimensional optical lattice, various geometries have been successfully created, such as square, triangular~\cite{triangle}, and Kagom{\'e}~\cite{Kagome} lattices.  
In the triangular and Kagom{\'e} lattices, the perfect nesting condition is not satisfied. 
Thus, the present study of a bipartite lattice satisfying the perfect nesting condition with the nesting wavevector $(\pi, \pi, \cdots)$ cannot be straightforwardly applied to them. 
Furthermore, when there is geometrical frustration, we expect interesting physics to emerge as a result of the interplay between the correlation effects and the frustration.

Finally, as a prospect, we discuss the relation between our theoretical results and the experiments. 
By considering hyperfine states to be the color degrees of freedom, we can realize the three-component lattice fermions with various atoms, such as $^6$Li and $^{173}$Yb. 
If Feshbach resonances are available, we can control the interactions by tuning the scattering lengths of the atoms including their magnitude, sign, and hyperfine dependences. 
For example, as shown in Fig. 2(c) in Ref. 15, the magnetic Feshbach resonance of $^6$Li allows us to control the color dependence of repulsive interactions.
In addition, by tuning the optical lattice potential, we can control the interaction strengths and the hopping integral.
In three-component $\rm ^6Li$ atoms, a very large loss rate was observed in the attractive region around a magnetic field of 500 G.\cite{Ottenstein2008} For attractively interacting three-component systems, it has been pointed out that loss-induced phase separation occurs.\cite{Priv,Titv} 
On the other hand, a very small three-body loss was observed in the repulsively interacting region under a magnetic field from 550 G to 600 G. 
We thus expect the $^6$Li atoms in optical lattices to be a possible candidate for observing  novel states such as a superfluid and a PMI. 
Another example for the three-component fermionic atoms is $\rm ^{173}Yb$ atoms with nuclear spin $I=5/2$.\cite{Taie2012,Taie2010} 
For $\rm ^{173}Yb$, where the electron spin is absent in the ground state, the hyperfine dependence of the scattering lengths is negligibly small; it provides the SU($N$) systems with $N=1,2, \cdots, 6$.\cite{Taie2010} 
The color-dependent interactions can be controlled by the optical Feshbach resonance.\cite{OpticalFB} 
The mixture of different isotopes, such as $\rm ^{173}Yb$ and $\rm ^{171}Yb$,\cite{Taie2010} are also a candidate: The three-component system can be realized by, for example, choosing two hyperfine states of $\rm ^{171}$Yb and one state of $\rm ^{173}$Yb.
In this system, the color-dependent interactions are naturally induced and can be controlled by the optical Feshbach resonance.
State-of-the-art experimental techniques, such as photo-association spectroscopy and rf-spectroscopy, can allow us to detect double occupancies and single-particle excitation spectra.\cite{JILA,Sugawa2011} 
As discussed above, these quantities exhibit key features for detecting phase transitions. 
We thus expect $\rm ^6Li$ atoms, $\rm ^{173}Yb$ atoms, and $\rm ^{173}Yb$-$\rm ^{171}Yb$ mixtures in optical lattices to be possible candidates for observing novel states such as the superfluid and the PMI. 
It is an interesting future direction to investigate the Mott transition, superfluidity, and staggered ordering of three-component fermionic atoms based on a realistic model Hamiltonian.

%

%
%
%

%%%%%%%%%%%%%%%%%%%%%%%%%%%%%%%%%%%%%%%%%%%%%%
\section*{Acknowledgments}
%This work was supported by JSPS KAKENHI (23540467).
This work was supported by Grant-in-Aid for Scientific Research (C) 
(No. 23540467) from the Japan Society for the Promotion of Science.
%%%%%%%%%%%%%%%%%%%%%%%%%%%%%%%%%%%%%%%%%%%%%%

\end{document}